\pdfoutput=1
\documentclass[conference]{IEEEtran}

\ifCLASSOPTIONcompsoc
  \usepackage[nocompress]{cite}
\else
  \usepackage{cite}
\fi

\usepackage{booktabs} 
\usepackage{tikz}
\usepackage{pgfplots}
\usepackage{amsmath}
\usepackage{amssymb}
\usepackage{amsfonts}
\usepackage{BOONDOX-cal} 
\usepackage{breakurl}
\usepackage{bbding}
\usepackage{listings}
\usepackage{url}
\usepackage{tabularx}
\usepackage{multirow}
\usepackage{makecell}
\usepackage{xr}
\usepackage{xcite}
\usepackage{fontenc}
\usepackage{graphicx}
\usepackage{balance}  
\usepackage{soul} 
\usepackage{times}
\usepackage{epsfig}
\usepackage{caption}
\usepackage{subcaption}
\usepackage{setspace}
\usepackage{indentfirst}
\usepackage[linesnumbered,ruled,noend]{algorithm2e}
\usepackage{algorithmicx}
\usepackage[noend]{algpseudocode}
\usepackage{pifont}
\usepackage{verbatim}
\usepackage{textcomp}

\newcommand{\nop}[1]{}

\newtheorem{definition}{\bf Definition}
\newtheorem{problem statement}{\bf Problem Statement}
\newtheorem{theorem}{\bf Theorem}
\newtheorem{lemma}{\bf Lemma}

\newcommand{\turboiso}{Turbo\textsubscript{ISO}}
\newcommand{\boostiso}{Boost\textsubscript{ISO}}
\newcommand{\quicksi}{QuickSI}

\newcommand{\qjoin}{\overset{T}{\Join}}  
\newcommand{\pjoin}{\overset{T}{\Join}}  
\newcommand{\pijoin}{\overset{T}{\Join}}

\newcommand{\iG}{\mathbb{G}}
\newcommand{\iV}{\mathbb{V}}
\newcommand{\iE}{\mathbb{E}}

\newcommand{\e}{\epsilon}
\newcommand{\n}{\mathcal{n}}

\newcommand{\mred}[1]{#1}
\newcommand{\mbred}[1]{#1}
\newcommand{\cmark}{\ding{51}}%
\newcommand{\xmark}{\ding{55}}%

\newcommand{\picfolder}{./pics/}

\newcommand{\timefolder}{./exp/time/}
\newcommand{\spacefolder}{./exp/space/}
\newcommand{\concurfolder}{./exp/concurrency/}

\newcommand{\djfolder}{./exp/decomp-join/}
\newcommand{\selectfolder}{./exp/selectivity/}

\newcommand{\myproof}[1]{
}
\newcommand{\optionshow}[2]{#1}

\renewcommand{\optionshow}[2]{#2}

\renewcommand{\picfolder}{./}

\renewcommand{\timefolder}{./}
\renewcommand{\spacefolder}{./}
\renewcommand{\concurfolder}{./}
\renewcommand{\djfolder}{./}
\renewcommand{\selectfolder}{./}

\begin{document}

\title{Time Constrained Continuous Subgraph Search over Streaming Graphs}


%
%

\author{%
{ {Youhuan Li${^\dag}$}, Lei Zou{${^\dag}$},  M. Tamer {\"O}zsu{${^\ddag}$},  Dongyan Zhao{${^{\dag}}$}}%
\\
\fontsize{10}{10}\selectfont\itshape $~^{\dag}$Peking University, China;
\fontsize{10}{10}\selectfont\itshape $~^{\ddag}$University of Waterloo,
Canada;
\\
\fontsize{10}{10}\selectfont\ttfamily\upshape $~^{\dag}$$\{$liyouhuan,zoulei,zhaody$\}$@pku.edu.cn, $~^{\ddag}$tamer.ozsu@uwaterloo.ca
\\}

\maketitle

\begin{abstract} 
The growing popularity of dynamic applications such as social networks provides a promising way to detect valuable information in real time. These applications create high-speed data that can be easily modeled as streaming graph.  Efficient analysis over these data is of great significance. In this paper, we study the subgraph (isomorphism) search over streaming graph data that obeys timing order constraints over the occurrence of edges in the stream. We propose a solution to efficiently answer subgraph search, introduce optimizations to greatly reduce the space cost, and design concurrency management to improve system throughput. Extensive experiments on real network traffic data and synthetic social streaming data confirms the efficiency and effectiveness of our solution.
\end{abstract}


\section{Introduction}\label{sec:introduction}
A recent development is the proliferation of high throughput, dynamic graph-structured data in many applications, such as social media streams and computer network traffic data. Efficient analysis of such streaming graph data is of great significance for tasks such as detecting anomalous events (e.g., in Twitter) and detecting adversarial activities in computer networks. Various types of queries over streaming graphs have been investigated, such as subgraph search, path computation, and triangle counting \cite{selectivityedbt15}. Among these, subgraph search is one of the most fundamental problems, especially subgraph isomorphism that provides an exact topological structure constraint for the search. 

In this paper, we study subgraph (isomorphism) search over streaming graph data that obeys timing order constraints over the occurrence of edges in the stream. Specifically, in a query graph, there exist some timing order constraints between different query edges specifying that one edge in the match is required to come before (i.e., have a smaller timestamp than) another one in the match. The timing aspect of streaming data is important for queries where sequential order between the query edges is significant. The following examples  demonstrate the usefulness of subgraph (isomorphism) search with timing order constraints over streaming graph data. 

{\bf Example 1. Cyber-attack pattern.}

\nop{
A command and control (C$\&$C) server is a centralized computer system that  issues commands to a botclient (zombie army) and receives reports back from the coopted computers, which  plays an important role in a botnet. C$\&$C servers can launch different kinds of attacks, such as ``information exfiltration'' and ``distributed denial of service (DDoS)''. As a core component of a botnet, C$\&$C servers always try various tricks to hide themselves from being tracked. Therefore, identifying, 
destroying or subverting C$\&$C servers are fundamental challenges in cybersecurity. Based on the attack behavior patterns over network traffic data, it is possible to identify these malware C$\&$C servers. A network packet message consists of source/target IPs as well as ports, protocols, message sizes, timestamps and so on. A packet message is represented as a directed edge between the source and the target and the whole network communication traffic data is a high-speed streaming graph. We can model an attack pattern as a query graph, reducing the problem to subgraph isomorphism of query graph over the data graph. 
} 

\begin{figure}[h!]
\centering
\resizebox{0.65\linewidth}{!}{
	\includegraphics{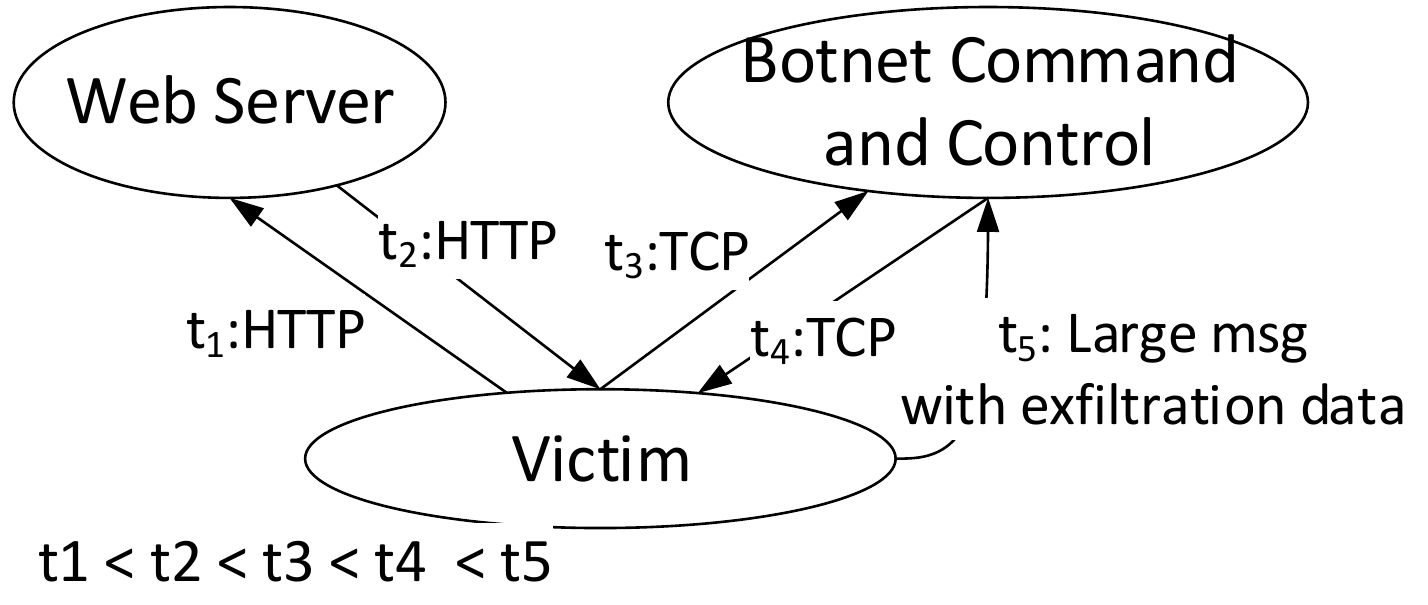}
}
\caption{Query example in Network Traffic (Taken from \protect\cite{selectivityedbt15})}
\label{fig:networktraffic}
\end{figure}

Figure \ref{fig:networktraffic} demonstrates the pipeline of the information exfiltration attack pattern. A victim browses a compromised website (at time $t_1$), which leads to downloading malware scripts (at time $t_2$) that establish communication with the botnet C$\&$C server (at times $t_3$ and $t_4$). The victim registers itself at the C$\&$C server at time $t_3$ and receives the command from the C$\&$C server at time $t_4$. Finally, the victim executes the command to send exfiltrated data back to C$\&$C server at time $t_5$. Obviously, the time points 
in the above example follow a strict timing order $t_1 < t_2< t_3 < t_4 < t_5$. Therefore, an attack pattern is modelled as a graph pattern ($Q$) as well as the timing order constraints over edges of $Q$. If we can locate the pattern (based on the subgraph isomorphism semantic) in the network traffic data, it is possible to identify the malware  C$\&$C Servers. US communications company Verizon has analyzed 100,000 security incidents from the past decade that reveal that 90$\%$ of the incidents fall into ten attack patterns \cite{verizon2016}, which can be described as graph patterns.

\begin{figure}[h!]
\centering
\resizebox{0.55\linewidth}{!}{
	\includegraphics{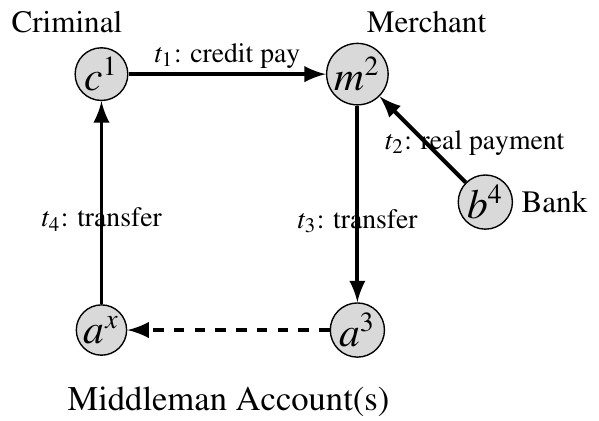}
}
\caption{Credit card  fraud in transactions (Taken from \protect \cite{cycledetection2018vldb})}
\label{fig:credit-card-fraud}
\end{figure}

{\bf Example 2. Credit-card-fraud pattern.}

\mbred{
Figure \ref{fig:credit-card-fraud} presents a credit card fraud example over a series transactions modeled by graph. A criminal tries to illegally cash out money by conducting a phony deal together with a merchant and a middleman. He first sets up a credit pay to the merchant ($t_1$); and when the merchant receives the real payment from the bank ($t_2$), he will transfer the money to a middleman ($t_3$) who will further transfer the money back to the criminal ($t_4$) to finish cashing out the money (Middleman may have more than one accounts forming transfer path).  Apparently,  this pattern where $t_1 <$ $t_2$ $<t_3$ $< t_4$  can be easily modeled as a query graph with timing order constraints. 
}


%

\subsection{Related Work}
Although subgraph search has been extensively studied in literature \cite{ullmann1976algorithm,vf2TKDBcordella2004sub,quicksi2008,turboiso,boostiso,sgi-closure2006he,sgi-krissinel2004common}, most of these works focus on static graphs. Ullman \cite{ullmann1976algorithm} proposes a well-known subgraph isomorphism algorithm that is based on a state-space search approach; Cordella et al. \cite{vf2TKDBcordella2004sub} propose the VF2 algorithm that employs several important pruning strategies when searching for targeted subgraphs. Shang et al. \cite{quicksi2008} employ filtering and verification strategy for subgraph isomorphism. They propose QI-sequence to  greatly reduce candidates from data graph before the verification phrase. Han et al. \cite{turboiso} transfer each query graph into a tree where they reduce duplicated subqueries to avoid redundant computation. They also utilize the tree to retrieve candidates from the data graph for further verification. Ren and Wang \cite{boostiso} define four vertex relationships over a query graph to reduce duplicate computation. 

The research on continuous query processing over high-speed streaming graph data is rather scarce. Fan et al. \cite{fan2013incremental} propose an incremental solution for subgraph isomorphism based on repeated search over dynamic graph data, which cannot utilize previously computed results when new data come from the stream since they do not maintain any partial result. To avoid the high overhead in building complicated index, there is some work on approximate solution to subgraph isomorphism. 
Chen et al. \cite{chen2010continuous} propose \emph{node-neighbor tree} data structure to search multiple graph streams; they relax the exact match requirement and their solution needs to conduct significant processing on the graph streams. Also, graph stream in \cite{chen2010continuous} is a sequence of small data graphs, which is not our focus.
Gao et al. \cite{gao2014continuous} study continuous subgraph search over a graph stream. They make specific assumption over their query and their solution cannot guarantee exact answers for subgraph isomorphism.
Song et al. \cite{eventpatternmatching14song} is the first work to impose timing order constraint in  streaming graphs, but the query semantics is based on \emph{graph simulation} rather than \emph{subgraph isomorphism}. The techniques for the former cannot be applied to the latter, since the semantics and, therefore, complexities are different. Furthermore, Song et al. perform post-processing to handle the timing constraints, i.e., finding all matches by ignoring the timing order constraints, and then filtering out the false positives based on the timing order constraints, which misses query optimization opportunities.
Choudhury et al. \cite{selectivityedbt15} consider subgraph (isomorphic) match over streaming graphs, but this work ignores timing order constraints. They propose a subgraph join tree (SJ-tree) to maintain some intermediate results, where the root contains answers for the query while the other nodes store partial matches. This approach suffers from large space usage due to maintaining results.

\optionshow{}{
To the best of our knowledge, this is the first work that investigates subgraph (isomorphism) matching over streaming graphs that take into account both structural and edge timing constraints. 
\optionshow{}{
Table \ref{tab:relatedwork} summarizes the differences between our work with the ones discussed above.
} 
} 

Due to the high speed of streaming graph data and the system's high-throughput requirement, a concurrent computing (i.e., multi-threaded) algorithm is desirable or even required. It is not trivial to extend a serial single-threaded algorithm to a concurrent one, as it is necessary to guarantee the consistency of concurrent execution over streaming graphs. 

\optionshow{}{
\begin{table}[!h]
\centering
\caption{Related work VS. Our Method}     
\label{tab:relatedwork}
\resizebox{\linewidth}{!}{
  \begin{small}
    \resizebox{\textwidth}{!}
    {
    \begin{tabular}{|c|c|c|c|}
    \hline
{\bfseries \makecell{Method}} & {\bfseries \makecell{Subgraph Isomorphism}} & {\bfseries \makecell{Timing Order}} & {\bfseries \makecell{Exact Solution}} \\ \hline
Our Method  									& \cmark & \cmark & \cmark \\ \hline
Choudhury et al. \cite{selectivityedbt15}   	& \cmark & \xmark & \cmark \\ \hline
Song et al. \cite{eventpatternmatching14song}   & \xmark & \cmark & \cmark \\ \hline
Gao et al. \cite{gao2014continuous}  			& \cmark & \xmark & \xmark \\ \hline
Chen et al. \cite{chen2010continuous}  			& \cmark & \xmark & \xmark \\ \hline
Fan et al. \cite{fan2013incremental}  			& \cmark & \xmark & \cmark \\ \hline
    \end{tabular}    }
\end{small} 
}
\end{table}
} 

\subsection{Our Solution and Contributions}
Our contributions are three-fold: (1) taking advantage of ``timing order constraints'' to reduce the search space, (2) compressing the space usage of intermediate results by designing a Trie-like data structure (called \emph{match-store tree}) and (3) proposing a concurrent computing framework with a fine-granularity locking strategy. The following is a summary of our methods and contributions:

\textbf{Reducing search space.} 
Considering the timing order constraints, we propose expansion list to avoid wasting time and space on \emph{discardable partial matches}. Informally, an intermediate result (partial match) $M$ is called ``discardable'' if $M$ cannot be extended to a complete match of query $Q$ no matter which edges would come in the future. Obviously, these should be pruned to improve the query performance. We define a query class, called \emph{timing connected-query} (TC-query for short--see Definition \ref{def:tcquery}) whose expansion list contains no discardable partial matches. We decompose a non-TC-query into a set of TC-queries and propose a two-step computing framework (Section \ref{sec:baseline}) . 

\textbf{Compressing space usage.} The materialization of intermediate results inevitably increases space cost, which raises an inherent challenge to handling massive-scale, high-speed streaming graphs. We propose a trie variant data structure, called \emph{match-store tree}, to maintain partial matches, which reduces both the space cost and the maintenance overhead without incurring extra data access burden (Section \ref{sec:compression}).  

\textbf{Improving system throughput.} 
Existing works do not consider concurrent execution of continuous queries over streaming graphs. For a high-speed graph stream, some edges may come at the same time. A naive solution is to process each edge one-by-one. In order to improve the throughput of the system, we propose to compute these edges concurrently. Concurrent computing may lead to conflicts and inconsistent results, which turns even more challenging when different partial matches are compressed together on their common parts. We design a fine-granularity locking technique to guarantee the consistency of the results (Section \ref{sec:concurrency}). 

\section{Problem Definition}\label{sec:problemdef}

\optionshow{
}
{We list frequently-used notations in Table \ref{tab:notations}.}

\begin{table}[!h]
\centering
\small
    \caption{\mbred{Frequently-used Notations}}     
    \label{tab:notations}
	\resizebox{\linewidth}{!}
	{
		    \begin{small}
    \begin{tabular}{|l|l|l|l|}
    \hline
        {\bfseries Notation} & \multicolumn{1} {c|} {\bfseries Definition and Description} \\
   \hline
        $\iG$ / $\iG_t$        & Streaming graph / Snapshot at time point $t$\\
   \hline
         $\iE_t$ / $\iV_t$     &  Edge/Vertex set of $\iG_t$\\
   \hline
        $Q$ / $V(Q)$ /   $E(Q)$      & Continuous query / Query vertex set / Query edge set \\
   \hline
        $\e_i$  /$\sigma_i$      & Query edge / Data edge at time $t_i$ \\
    \hline
        $g$  & A subgraph of some snapshot\\
    \hline
        $\overrightarrow{uv}$  & The directed  edge from vertex $u$ to $v$\\
    \hline
        $W$    & Time window $W$  \\
    \hline
        $\prec$    & Timing order over query edges    \\
    \hline
        $Preq(\e_i)$ & Prerequisite subquery of query edge $\e_i$   \\
    \hline
        $P_i$   & TC-subquery   \\  
    \hline
        $L_i (i>0)$   & Expansion list for TC-subquery $P_i$   \\
    \hline
        $L_0$   & Expansion list for joining matches of all TC-subqueries: \{$P_1$, $P_2$,...,$P_k$\}   \\      
    \hline
        $L_i^j$   & The $j$-th item in expansion list $L_i$  \\
    \hline
        $\Omega(q)$   & Matches of subquery $q$  \\    
    \hline
        $\Delta(q)$   & New matches of subquery $q$  \\
    \hline
        $D$   & A decomposition (set of TC-subqueries) of query $Q$  \\
    \hline
        $Ins(\sigma)$   & Insertion for incoming edge $\sigma$   \\
    \hline
        $Del(\sigma)$   & Deletion for expired edge $\sigma$   \\
    \hline
        $\n$ / $\n_i^j$    & A node in a MS-tree / The $j$-th node in the MS-tree for $L_i$ \\
    \hline
        $TCsub(Q)$    & The set of all TC-subqueries of query $Q$ \\
    \hline
    \end{tabular}
    \end{small}

	}
\end{table}

\nop{ 
\begin{definition}[Streaming Graph]
A streaming graph $\iG$ is an unbounded time-evolving sequence of directed edges $\{\sigma_1,...,\sigma_{\infty}\}$ in which each $\sigma_i$ arrives at a particular time $t_i$ ($t_i \leq t_j$ when $0<i<j$). This time point is also referred to as the timestamp of $\sigma_i$. Each edge $\sigma_i$ has two labelled vertices and two edges are connected if and only if they share one common endpoint. 
\end{definition}
} 

\begin{definition}[Streaming Graph] \label{def:graphstream}
A streaming graph $\iG$ is \mbred{a constantly growing sequence} of directed edges $\{\sigma_1, \sigma_2, ...\sigma_x\}$ in which each $\sigma_i$ arrives at a particular time $t_i$ ($t_i < t_j$ when $0<i<j$). This time point is also referred to as the timestamp of $\sigma_i$. Each edge $\sigma_i$ has two labelled vertices and two edges are connected if and only if they share one common endpoint. 
\end{definition}


For simplicity of presentation, we only consider vertex-labelled graphs and ignore edge labels, although handling the
more general case is not more complicated. For example, since vertex labels and edge labels are from two different label sets, we can introduce an imaginary vertex to represent an edge of interest and assign the edge label to the new imaginary vertex.

\begin{figure}[h!]
\centering
\resizebox{0.80\linewidth}{!}{
	\includegraphics{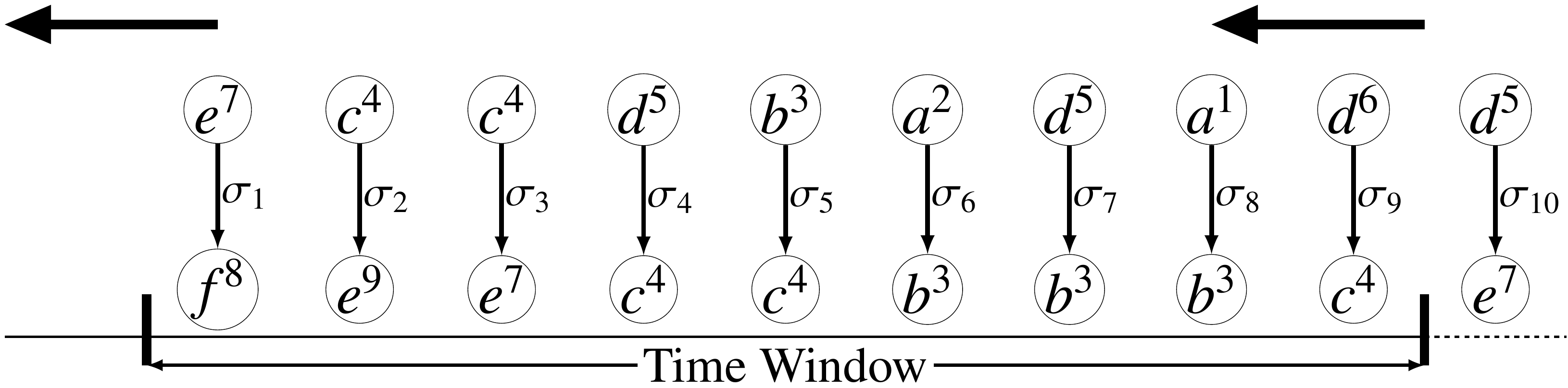}
}
\caption{Graph stream $\iG$ under time window of size $9$}
\label{fig:gstream}
\end{figure}

An example of a streaming graph $\iG$ is shown in Figure \ref{fig:gstream}. Note that edge $\sigma_1$ has two endpoints $e^7$ and $f^8$, where `$e$' and `$f$' are vertex labels and the superscripts are vertex IDs that we introduce to distinguish two vertices with the same label.


\nop{ 
In this paper, we use the \emph{time-based sliding window model}, where a sliding window $W$ defines a timespan with fixed duration $|W|$. If the current time is $t_i$, the time window $W$ defines the timespan ($t_i-|W|, t_i$]. Obviously, all edges that occur in this time window form a consecutive block over the edge sequence $\{\sigma_1,...,\sigma_{\infty}\}$. At time $t_{i+1}$, some edges may expire and some new edges will be appended according to the updated time window $W$.
}

In this paper, we use the \emph{time-based sliding window model}, where a sliding window $W$ defines a timespan with fixed duration $|W|$. 
If the current time is $t_i$, the time window $W$ defines the timespan ($t_i-|W|, t_i$]. Obviously, all edges that occur in this time window form a consecutive block over the edge sequence and as time window $W$ slides, some edges may expire and some new edges may arrive.

\begin{figure}[h!]
\newcommand{\mywidth}{0.3\linewidth}
\newcommand{\mylinewidth}{\linewidth}
\centering
	\begin{subfigure}[t]{\mywidth}
		\centering
		\resizebox{\mylinewidth}{!}
		{
			\includegraphics{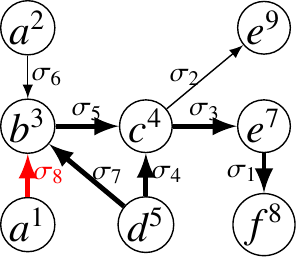}
		}
		\caption{$t = 8$}
		\label{fig:stream8}
	\end{subfigure}
	\begin{subfigure}[t]{\mywidth}
		\centering
		\resizebox{\mylinewidth}{!}
		{
			\includegraphics{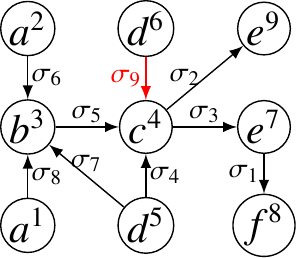}
		}
		\caption{$t = 9$}
		\label{fig:stream9}
	\end{subfigure}
	\begin{subfigure}[t]{\mywidth}
		\centering
		\resizebox{\mylinewidth}{!}
		{
			\includegraphics{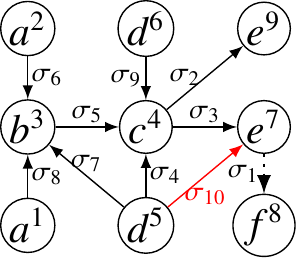}
		}
		\caption{$t = 10$}
		\label{fig:stream10}
	\end{subfigure}
	\caption{Graph stream under time window $W$ of size $9$}
	\label{fig:wgstream}
\end{figure}  

\begin{figure}[h!]
\newcommand{\mywidth}{0.35\linewidth}
\centering
	\begin{subfigure}[t]{\mywidth}
		\centering
		\resizebox{\linewidth}{!}
		{
			\includegraphics{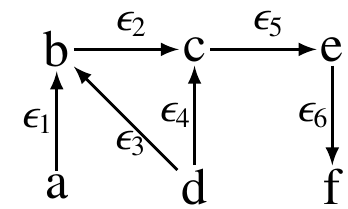}
		}
		\caption{query graph}
		\label{fig:qgraph}
	\end{subfigure}
	\begin{subfigure}[t]{\mywidth}
		\centering
		\resizebox{0.8\linewidth}{!}
		{
			\includegraphics{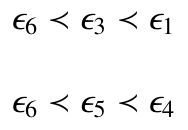}
		}
		\caption{timing order}
		\label{fig:qtiming}
	\end{subfigure}
	\caption{Running example query $Q$}
	\label{fig:examplequery}
\end{figure}

\begin{definition}[A Snapshot of a Streaming Graph]  \label{def:snapshot}
Given a streaming graph $\iG$ and a time window $W$ \mbred{at current time point $t$, the current snapshot of $\iG$ is a graph $\iG_t=$ $(\iV_t, \iE_t)$ where $\iE_t$ is the set of edges that occur in $W$ and $\iV_t$ is the set of vertices adjacent to edges in $\iE_t$, namely:}
\[
\iE_t = \{\sigma_i | t_i \in (t-|W|, t]\},\iV _t = 
\{u | \overrightarrow{uv} \in \iE_t 
\lor \overrightarrow{vu} \in \iE_t \}
\]
\end{definition}

The snapshots of graph stream $\iG$ at time points $t=8, 9, 10$ for $|W|=9$ are given in Figure \ref{fig:wgstream}. Note that at timestamp $t={10}$, edge $\sigma_{1}$ expires since the time point of $\sigma_{1}$ is 1 and the timespan of time window $W$ is $(1,10]$. The expired edges are denoted with dotted edges in Figure \ref{fig:stream10} while newly added edges are in red.

\begin{definition}[Query Graph]
\label{def:query}
A query graph is a four-tuple $Q$ = ($V(Q)$, $E(Q)$, $L$, $\prec)$, where $V(Q)$ is a set of vertices in $Q$, $E(Q)$ is a set of directed edges, $L$ is a function that assigns a label for each vertex in $V(Q)$, and  $\prec$  is a strict partial order relation over $E(Q)$, called the \emph{timing order}. For $\e_i$, $\e_j$ $\in E(Q)$, $\e_i$ $\prec \e_j$ means that in a match $g$ for $Q$ where $\sigma_i$ matches $\e_i$ and $\sigma_j$ matches $\e_j$ ($\sigma_i$, $\sigma_j$ $\in g$), timestamp of $\sigma_i$ should be less than that of $\sigma_j$. 
\end{definition}
An example of query graph $Q$ is presented in Figure \ref{fig:examplequery}. Any subgraph in the result must conform to the constraints on both structure and timing orders. For example, in query $Q$, $\e_1 \prec \e_2$ ($\e_1,\e_2 \in E(Q)$) means that edges matching $\e_1$ should arrive before edges matching $\e_2$ in subgraph matches of $Q$ over the snapshot (see Definition \ref{def:subgraphmatch}) in the current time window.

\begin{definition}[Time-Constrained Match]\label{def:subgraphmatch} For a query $Q$ and a subgraph $g$ in current snapshot, $g$ is a \emph{time-constrained match} of $Q$ if only if there exists a bijective function $F$ from $V(Q)$ to $V(g)$ such that the following conditions hold: 
\begin{enumerate}
	\item \textbf{Structure Constraint (Isomorphism)} 
		\begin{itemize}
		\item 
			$\forall u \in V(Q),L(u) = L(F(u))$. 
		\item 
			$\overrightarrow {uv}  \in E(Q) \Leftrightarrow \overrightarrow {F(u)F(v)}  \in E(g)$.
		\end{itemize}
	\item   \textbf{Timing Order Constraint} \\
		For any two edges $(\overrightarrow{u^{i_1} u^{i_2}})$, $(\overrightarrow{u^{j_1} u^{j_2}})$ $\in E(Q)$:
		\[
			(\overrightarrow{u^{i_1} u^{i_2}})  \prec (\overrightarrow{u^{j_1} u^{j_2}}) \Rightarrow  \overrightarrow{F(u^{i_1}) F(u^{i_2})} \prec \overrightarrow{F(u^{j_1}) F(u^{j_2})}
		\]
\end{enumerate}
\end{definition}

Hence, the problem in this paper is to find all \emph{time-constrained matches} of given query $Q$ over each snapshot of graph stream $\iG$ with window $W$.  For simplicity, when the context is clear, we always use ``match'' to mean ``time-constrained match''. 

For example, the subgraph $g$ induced by edges $\sigma_1$, $\sigma_3$, $\sigma_4$, $\sigma_5$, $\sigma_7$ and $\sigma_8$ in Figure \ref{fig:stream8}  (highlighted by bold line) is not only isomorphic to query $Q$ but also conforms to the timing order constraints defined in Figure \ref{fig:qtiming}. Thus, $g$ is a match of query $Q$ over stream $\iG$ at time point $t=8$.  At time point $t=10$, with the deletion of edge $\sigma_1$,  $g$ expires. 

\begin{theorem}\label{theorem:nphard}
\mbred{Subgraph isomorphism can be reduced to the proposed problem in polynomial time and therefore, the proposed problem is NP-hard.}
\end{theorem}
\optionshow{}{
\begin{proof}
Consider an arbitrary subgraph isomorphism problem: given two graphs $G$ and $g$,  answering whether these is a subgraph $g^{\prime}$ of $G$ that is isomorphic to $g$. Let's reduce this problem into a time constrained continuous subgraph isomorphism over streaming graph. Assuming that there are $m$ edges in $G$: \{$\sigma_1$, $\sigma_2$, ..., $\sigma_{m}$\},  we  transform $G$ into a streaming graph $\iG$ by randomly assigning timestamp $t_i$ to each edge $\sigma_i$ such that $0< t_i$ $< t_j$ if $i<j$. We set a time window $W$ of size $(t_m - t_1)$ and initial timespan $(0, t_1]$. Let $g$ be a continuous query graph of timing order $\prec$ $=\emptyset$. In this way, let's apply our solution over the streaming graph $\iG$ of time window $W$ with query $g$, if there are answers for $g$ at time $t_m$, then there must be subgraph $g^{\prime}$ of $G$ that is isomorphic to $g$. Apparently, it costs only polynomial time to finish reducing a subgraph isomorphism problem to the proposed one. Hence, the proposed problem is NP-hard. 
\end{proof}
}

\section{A Baseline Method} \label{sec:baseline}
We propose a baseline solution that utilizes the timing order in reducing the search space. We first define a class of queries  (timing-connected query)  and the corresponding evaluation in Section \ref{sec:baseline:tcq}; we then discuss how to answer an arbitrary query in Section \ref{sec:baseline:nontcq}.

\subsection{Timing-Connected Query}\label{sec:baseline:tcq}
\subsubsection{Intuition} \label{sec:tcq:intuition}
A naive solution to executing a query $Q$ with timing order is to run a classical subgraph isomorphism algorithm (such as \quicksi \cite{quicksi2008}, \turboiso \cite{turboiso}, \boostiso \cite{boostiso}) on each snapshot $\iG_i$ $(i=1,...,\infty)$ to first check the structure constraint followed by a check of the timing order constraint among the matches. Obviously, this is quite expensive. A better approach is to identify the subgraph $\Delta(\iG_i)$ of $\iG_i$ that is affected by the updated edge (insertion/deletion) and then conduct subgraph isomorphism algorithm over $\Delta(\iG_i)$  instead of the whole snapshot $\iG_i$. While, if the query diameter is $d$, then $\Delta(\iG_i)$ is  the subgraph induced by all vertices that is $d$-hop reachable to/from the adjacent vertices of the updated edge \cite{fan2013incremental}. Hence, the size of $\Delta(\iG_i)$ could be huge if query diameter is large which results in the inefficiency of the computation.

However, an incoming/expired edge causes only a minor change between two consecutive snapshots $\iG_i$ and $\iG_{i-1}$; thus, it is wasteful to re-run the subgraph isomorphism algorithm from scratch on each snapshot. Therefore, we maintain \emph{partial matches} of subqueries in the previous snapshots. Specifically, we only need to check whether there exist some partial matches (in the previous snapshots) that can join with an incoming edge $\sigma$ to form new matches of query $Q$ in the new snapshot $\iG_{i}$. Similarly, we can delete all (partial) matches containing the expired edges at the new timestamp. 
For example, consider the query graph $Q$  in Figure \ref{fig:examplequery}. Assume that an incoming edge $\sigma$ matches $\e_1$ at time point $t_i$. If we save all partial matches for subquery $Q\backslash$$\{\e_1\}$, i.e., the subquery induced by edges $\{\e_2$, $\e_3$, $\e_4$, $\e_5$, $\e_6\}$, at the previous time point $t_{i-1}$ (i.e., $\iG_{i-1})$, we only need to join $\sigma$ with these partial matches to find new subgraph matches of query $Q$. 


Although materializing partial matches can accelerate continuous subgraph query, it is inevitable to introduce much maintenance overhead. For example, 
in SJ-tree \cite{selectivityedbt15}, each new coming edge $\sigma$ requires updating the partial matches. In this section, we propose pruning \emph{discardable} edges (see Definition \ref{def:discardable}) by considering the timing order in the query graph. 

\nop{
Generally, if an edge $\sigma$ is discardable, it means that $\sigma$ never contributes to any match of query $Q$ no matter what edges of the streaming graph arrive in the future.  
}

\begin{definition}[Discardable Edge]\label{def:discardable}
\mred{For a streaming graph $\iG$ and a query graph $Q$,  an incoming edge $\sigma$ is called a \emph{discardable edge} if $\sigma$ cannot be included in a complete match of $Q$, no matter what edges arrive in the future.  }
\end{definition} 






\mbred{To better understand discardable edge, recall the streaming graph} $\iG$ in Figure \ref{fig:gstream}. At time $t_6$, an incoming edge $\sigma_6$ (only matching $\e_1$) is added to the current time window. Consider the timing order constraints of query $Q$ in Figure \ref{fig:examplequery}, which requires that edges matching $\e_3$ should come before ones matching $\e_1$. However, there is no edge matching $\e_3$ before $t_6$ in $\iG$. Therefore, it is impossible to generate a complete match (of $Q$) consisting of edge $\sigma_6$ (matching $\e_1$) no matter which edges come in the future. Thus, $\sigma_6$ is a \emph{discardable edge} that can be filtered out safely. 
We design an effective solution to determine if an  incoming edge $\sigma$ is discardable. Before presenting our approach, we introduce an important definition.

\begin{definition}[Prerequisite Edge/Prerequisite Subquery]
\label{def:prerequisite}
Given an edge $\e$ in query graph $Q$, a set of \emph{prerequisite edges} of $\e$ (denoted as $Preq(\e))$ are defined as follows:
\[
Preq(\e) = \{ \e^{\prime} | \e^{\prime}  \prec \e \}  \cup \{ \e \} 
\]
where `$\prec$' denotes the timing order constraint as in Definition \ref{def:query}.
The subquery of $Q$ induced by edges in $Preq(\e )$ is called a \emph{prerequisite subquery} of $\e$ in query $Q$. 
\end{definition}

\vspace{-0.1in}
\begin{figure}[h!]
\centering
	\begin{subfigure}[t]{0.26\linewidth}
		\centering
		\resizebox{\linewidth}{!}
		{
			\includegraphics{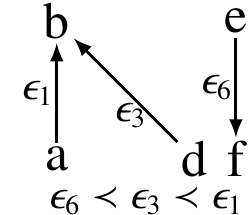}
		}
		\caption{$Preq(\e_1)$}
		\label{fig:preqe1}
	\end{subfigure}
	\hspace{0.15in}
	\begin{subfigure}[t]{0.25\linewidth}
		\centering
		\resizebox{0.95\linewidth}{!}
		{
			\includegraphics{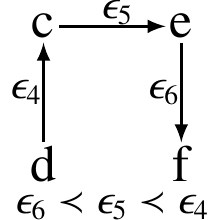}
		}
		\caption{$Preq(\e_4)$}
		\label{fig:preqe4}
	\end{subfigure}
	\caption{Example of prerequisite subquery}
	\label{fig:example:preq}
\end{figure} 
\vspace{-0.1in}

Consider two edges $\e_1$ and $\e_4$ in query $Q$ in Figure \ref{fig:examplequery}. Prerequisite subqueries $Preq(\e_1)$ and $Preq(\e_4)$ are both illustrated in Figure \ref{fig:example:preq}.  The following lemma states the necessary and sufficient condition to determine whether an edge $\sigma$ in  streaming graph $\iG$ is \emph{discardable}\optionshow{(All proofs of lemmas and theorems are presented in the full version of this paper \cite{fulltimingsubg})}{}.

\begin{lemma}\label{lem:condition:promising}
An incoming edge $\sigma$ at time $t_i$ is \emph{NOT} \emph{discardable} if and only if, at the current snapshot $\iG_i$, there exists at least one query edge $\e$ ($\in Q$) such that (1) the prerequisite subquery $Preq(\epsilon)$ has at least one match $g$ (subgraph of $\iG_i$) containing $\sigma$; and (2) $\sigma$ matches $\e$ in the match relation between $g$ and $Preq(\e)$. Otherwise, $\sigma$ is \emph{discardable}. 
\end{lemma}
\optionshow{}{
\begin{proof}
If $\sigma$ is not discardable and $g^{\prime}$ is the match in the future that includes $\sigma$, then there must be a query edge $\e$ that $\sigma$ matches. Also, since $Preq(\e)$ is a subquery of query $Q$, we can always find a subgraph $g$ $\subseteq$ $g'$ such that $g$ matches $Preq(\e)$ and $\sigma$ $\in g$. On the contrary, if $\sigma$ matches $\e$ and there is a subgraph $g$ where $\sigma$ $\in g$ and $g$ matches $Preq(\e)$, then all edges that are required to be before $\sigma$ have been in the time window and it is possible that there will be a series edges in the future that extend $g$ into a match of $Q$, indicating that $\sigma$ is not discardable. 
\end{proof}
}

Lemma \ref{lem:condition:promising}  can be used to verify whether or not an incoming edge $\sigma$ is discardable. The straightforward way requires checking subgraph isomorphism between $Preq(\e)$ and $\iG_i$ in each snapshot, which is quite expensive. First, $Preq(\e)$ may not be connected, even though query $Q$ is connected. For example, $Preq(\e_1)$ (in Figure \ref{fig:preqe1}) is disconnected. Computing subgraph isomorphism for disconnected queries will cause a Cartesian product among candidate intermediate results leading to lots of computation and huge space cost. Second, some different prerequisite subqueries may share common substructures, leading to common computation for different prerequisite subqueries. It is inefficient to compute subgraph isomorphism from scratch for each incoming edge. 

For certain types of queries that we call \emph{timing-connected query} (Definition \ref{def:tcquery}), it is easy to determine if an edge $\sigma$ in streaming graph $\iG$ is discardable. Therefore, we first focus on these queries for which we design an efficient query evaluation algorithm. We discuss non-TC-queries in Section \ref{sec:baseline:nontcq}. 

For ease of presentation, we introduce the following concepts that will be used when illustrating our algorithm.  Consider a query $Q$ and two subqueries: $Q^1$, $Q^2$, assume that $g_1$  ($g_2$) is a time-constrained match of $Q^1$ ($Q^2$) in the current snapshot. Let $F_1$ and $F_2$ denote the \emph{matching functions} (Definition \ref{def:subgraphmatch}) from $V(Q^1)$ and $V(Q^2)$ to $V(g_1)$ and $V(g_2)$, respectively. We say that $g_1$ is \emph{compatible} with $g_2$ (denoted as $g_1 \sim$ $g_2$) W.R.T $Q^1$ and $Q^2$ if and only if $g_1$ $\cup$ $g_2$ is a time-constrained match of $Q^1$ $\cup$ $Q^2$ on bijective match function $F_1$ $\cup$ $F_2$.  Furthermore, let $\Omega(Q^1)$ and $\Omega(Q^2)$ denote the set of matches of $Q^1$ and $Q^2$ in current snapshot, respectively. We define a new join operation over $\Omega(Q^1)$ and $\Omega(Q^2)$, denoted as $\Omega(Q^1)$ $\qjoin \Omega(Q^2)$, as follows:
\[
	\Omega(Q^1)\qjoin \Omega(Q^2) = \{ g_1 \cup g_2 |  g_1\in \Omega(Q^1) \sim g_2\in \Omega(Q^2) \}
\] 
\mbred{Note that when $g_1$ $\sim$ $g_2$ and $Q^1$ $\cap$ $Q^2$ $\neq \emptyset$, $F_1$ and $F_2$ will never map the same query vertex to different data vertices since we require $F_1\cup F_2$ to be a bijective function}.


\subsubsection{TC-query} \label{sec:tcq:definiton}

\begin{definition}[Prefix-connected Sequence] \label{def:pre:conn:seq}
Given a query $Q$ of $k$ edges, a \emph{prefix-connected sequence} of $Q$ is a permutation of all edges in $Q$: $\{\e_1$, $\e_2$...,$\e_k\}$ such that $\forall j \in [1, k]$, the subquery induced by the first $j$ edges in $\{\e_1\}\cup...\cup \{\e_j\}$ is always \emph{weakly connected}.
\end{definition}

\begin{definition}[Timing-connected Query] \label{def:tcquery}
A query $Q$ is called a timing-connected query (\emph{TC-query} for short) if there exists a prefix-connected sequence $\{\e_1$, $\e_2$...,$\e_k\}$ of $Q$ such that $\forall j \in [1, k-1]$, $\e_j \prec \e_{j+1}$. In this case, we call the sequence $\{\e_1$,...,$\e_k\}$ the \textbf{timing sequence} of TC-query $Q$.
\end{definition}

Recall the running example $Q$ in Figure \ref{fig:examplequery}, which is not a TC-query. However, the subquery induced by edges  $\{\e_6$, $\e_5$, $\e_4\}$ is a TC-query, since $\e_6 \prec \e_{5} \prec \e_4$ and  $\{\e_6\}$, $\{\e_6$, $\e_5\}$ and $\{\e_6$, $\e_5$, $\e_4\}$ are all connected. 


Given a TC-query $Q$ with timing sequence  $\{\e_1$,...,$\e_k\}$,
the prerequisite subquery $Preq(\e_j)$ is exactly the subquery induced by the first $j$ edges in $\{\e_1$, $\e_2$,...,$\e_j\}$ $(j \in [1,k])$. $Preq(\e_{j+1})= Preq(\e_{j}) \cup \{e_{j+1}\}$ and $\Omega(Preq(\e_{j+1}))=\Omega(Preq(\e_{j})) \pjoin \Omega(\e_{j+1}) $, where  $\Omega(Preq(\e_{j+1}))$ denotes matches for prerequisite subquery $Preq(\e_{j+1})$, $\Omega(\e_{j+1})$ denotes the matching edges for $\e_{j+1}$.

\subsubsection{TC-query Evaluation} \label{sec:tcq:evaluation}
We propose an effective data structure, called \emph{expansion list}, to evaluate a TC-query $Q$. An expansion list for TC-query (1) can efficiently determine whether or not an incoming edge is discardable, and (2) can be efficiently maintained (which guarantees the efficient maintenance of the answers for TC-query $Q$). 

\begin{definition}[Expansion List] \label{def:expansion:list}
Given a TC-query $Q$ with timing sequence $\{\e_1$, $\e_2$,...,$\e_k\}$, an expansion list $L=$ $\{L^1$,$L^2$,...,$L^k\}$ over $Q$ is defined as follows:
\begin{enumerate}
\setlength{\itemsep}{1pt}
\item 
Each item $L^i$ corresponds to $\bigcup_{j=1}^{i}(\e_j)$, i.e., $Preq(\e_i)$.
\item 
Each item $L^i$ records $\Omega(\bigcup_{j=1}^{i}(\e_j))$, i.e., a set of partial matches (in the current snapshot) of prerequisite subquery $Preq(\e_i)$ $(i \in [1,k])$. We also use $\Omega(L^i)$ to denote the set of partial matches in $L^i$.
\end{enumerate}
\end{definition}

Note that each item $L^j$ corresponds to a distinct subquery $Preq(\e_j)$ and we may use the corresponding subquery to denote an item when the context is clear.

\begin{figure}[h!]
\centering
    \begin{subfigure}[t]{0.25\linewidth}
	    \centering
	    \resizebox{\linewidth}{!}
	    {
	        \includegraphics{\picfolder preqe4}
	    }
	    \caption{TC-query}
	    \label{fig:telist:tcq}
    \end{subfigure}
    \begin{subfigure}[t]{0.70\linewidth}
        \centering
        \resizebox{\linewidth}{!}
        {
            \includegraphics{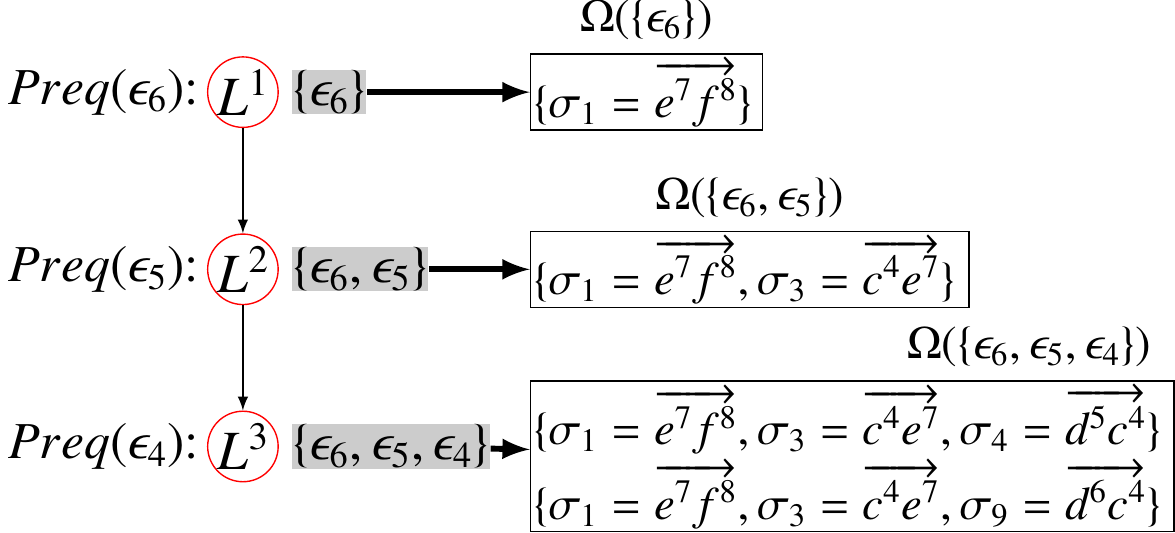}
        }
        \caption{Expansion list}
        \label{fig:telist:store}
    \end{subfigure}
\caption{A TC-query $\{\e_6$, $\e_5$, $\e_4\}$ and  timing expansion list}
\label{fig:timing:expansion:list}
\end{figure}

The shaded nodes in Figure \ref{fig:timing:expansion:list} illustrate the prerequisite subqueries for a TC-query  with timing sequence $\{\e_6$, $\e_5$, $\e_4\}$. Since each node corresponds to a subquery $Preq(\e_i)$, we also record the matches of $Preq(\e_i)$, as shown in Figure \ref{fig:timing:expansion:list}. The last item stores matches of the TC-query in the current snapshot. 

Maintaining the expansion list requires updating (partial) matches associated with each item in the expansion list. An incoming edge may result in insertion of new (partial) matches into the expansion list while an expired edge may lead to deletion of partial matches containing the expired one. We will discuss these two cases separately.

\textbf{Case 1: New edge arrival.}
For an incoming edge $\sigma$, Theorem \ref{theorem:tcquerydetermine} tells us which (partial) matches associated with the expansion list should be updated.

\begin{theorem}\label{theorem:tcquerydetermine} Given a TC-query $Q$ with the timing sequence $\{\e_1$, $\e_2$ ,..., $\e_k\}$ and the corresponding expansion list $L=$ $\{L^1$, $L^2$,...,$L^k\}$. If an incoming edge $\sigma$ matches query edge $\e_i$ in the current time window, then only the (partial) matches of $L^i$ $(Preq(\e_{i}))$ should be updated in the current snapshot.   
\begin{enumerate}
\setlength{\itemsep}{1pt}
\item If $i=1$, $\sigma$ should be inserted into $L^1$ as a new match of $Preq(\e_{1})$ since $Preq(\e_{1})$ $=\{\e_1\}$.
\item If $i \neq 1 \wedge  \Omega(L^{i-1}) \pjoin \{\sigma\} \neq \emptyset$, then $\Omega(L^{i-1}) \pjoin \{\sigma\}$ should be inserted into $L^i$ as new matches of $Preq(\e_{i})$, where $\Omega(L^{i-1})$ is the set of partial matches in $L^{i-1}$.
\end{enumerate}
\end{theorem}
\optionshow{}{
\begin{proof}
Assume that incoming edge $\sigma$ (matching $\e_i$) causes new partial matches that should be inserted into $L^{i'}$. If $i' < i$, then $\e_i$ $\notin Preq(\e_{i'})$, which means there will be no partial matches in $L^{i'}$ containing $\sigma$. However if $i' > i$, assuming that $g=$ $\{\sigma_1$,$\sigma_2$,...,$\sigma_{i'}\}$ is a new match to be inserted into $L^{i'}$ where $\sigma_{i^{\prime\prime}}$ matches $\e_{i^{\prime\prime}}$ ($1\leq i^{\prime\prime}\leq i^{\prime}$ and $\sigma=\sigma_{i}$), then $\sigma$ ($\sigma_{i}$) has smaller timestamp than that of $\sigma_{i^{\prime}}$ because $\e_i$ $\prec \e_{i^{\prime}}$, which is impossible since $\sigma$ is the incoming edge of largest timestamp in the current window. Thus, $i^{\prime}$ must be $i$.
\end{proof}
}

Hence, for a TC-query $Q$ $=\{\e_1$, $\e_2$...,$\e_k\}$ and the corresponding expansion list $L =$ $\{L^1$,$L^2$,...,$L^k\}$, the maintenance of $L$ for an incoming edge $\sigma$ can be done as follows:

\begin{enumerate}
\item 
	if $\sigma$ matches no query edge, discard $\sigma$;
\item
	if $\sigma$ matches $\e_1$, then add $\sigma$ into $L^1$;
\item
	if $\sigma$ matches $\e_i$ ($i>1$), then compute $\Omega(L^{i-1}) \pjoin \{\sigma\}$. If the join result is not empty, add all resulting (partial) matches (of $Preq(\e_{i})$) into $L^{i}$.
\end{enumerate}

\begin{theorem} \label{theorem:filteringtime}
\mbred{
Given a TC-query $Q$ $=\{\e_1$, $\e_2$...,$\e_k\}$ and the corresponding expansion list $L =$ $\{L^1$,$L^2$,...,$L^k\}$, for an incoming edge $\sigma$ that matches $\e_i$,  the time to determine whether $\sigma$ is discardable (to be filtered) or not is $O(|L^{i-1}|)$, which is linear to the number of partial matches in $L^{i-1}$.}
\end{theorem}
\optionshow{}{
\begin{proof}
Obviously, the main cost for determining whether $\sigma$ is discardable (to be filtered) or not lies in the join between \{$\sigma$\} and $\Omega(L^{i-1})$, which costs $O(|L^{i-1}|)$ time. $L^{i-1}$ contains all matches of subquery \{$\e_1$, $\e_2$...,$\e_{i-1}$\} and its size range from $0$ to $|\iE_t|^{i-1}$.
\end{proof}
}

The above process is codified in Lines \ref{code:b:tcqeva}-\ref{code:e:tcqeva} of Algorithm \ref{alg:insert}. 
Note that an incoming edge $\sigma$ may match multiple query edges; the above process is repeated for each matching edge $\e$. New matches that are inserted into the last item of the expansion list are exactly the new matches of TC-query $Q$. 
\optionshow{}{
For example, consider the TC-query $Q=$$\{\e_6$, $\e_5$, $\e_4\}$ in Figure \ref{fig:telist:tcq} and the streaming graph $\iG$ in Figure \ref{fig:gstream}. At time $t=2$, when the edge $\sigma_2$ (matching $\e_5$) arrives, $\Omega(Preq(\e_6))=\{\sigma_1\} \pjoin \{\sigma_2\}=\emptyset$, so there is no new match that is inserted into $\Omega(Preq(\e_5))$.
} 

\begin{algorithm}[h!]
\small
\caption{INSERT($\sigma$)}
 \label{alg:insert}
\KwIn{$\sigma$: incoming edge to be inserted}
\KwIn{$L_i$ $=\{L_i^1$, $L_i^2$,...,$L_i^{|Q^i|}\}$: the expansion list for $Q^i$}
\KwIn{$L_0$ $=\{L_0^1$, $L_0^2$,...,$L_0^{k}\}$: the expansion list over $\{Q^1$,$Q^2$,...,$Q^k\}$}

\For{each query edge $\e$ that $\sigma$ matches} 
{
	\label{code:b:tcqeva}
	Assume that $\e$ is the $j$-th edge in TC-subquery $Q^i$. \\
	\If{$j==1$}
	{ 
	   Insert $\sigma$ into $L_i^j$  \label{code:lminsert}
	}
	\Else
	{ 
		Let $\Delta(\e)=\{\sigma\}$ \label{code:b:join}\\
	    READ($L_i^{j-1}$)  {\textbf{// Read partial matches in $L_i^{j-1}$} }  \\    
		$\Delta(L_i^j)=\Delta(\e) \pijoin \Omega(L_i^{j-1})$\\
		\If{$\Delta(L_i^j) \neq \emptyset$}
		{
	        INSERT($\Delta(L_i^j)$, $L_i^j$) \textbf{// Insert $\Delta(L_i^j)$ into $L_i^j$} \label{code:e:join} \label{code:ins1}   \label{code:e:tcqeva}
	    }
	}
	\If{$j=|L_i|$ AND $\Delta(L_i^j) \neq \emptyset$}
	{
		\If{$i=1$}
		{
			Let $\Delta(L_{0}^{i}) = \Delta(L_i^j)$
		}
		\Else
		{\label{code:begin:ilargerone}
			READ($L_0^{i-1}$) \textbf{// Read partial matches in $L_0^{i-1}$}\\       
			$\Delta(L_0^i)=\Delta(L_i^{j}) \qjoin \Omega(L_0^{i-1})$\label{code:l0:firstjoin}\\  
			INSERT($\Delta(L_0^i)$, $L_0^i$) \textbf{// Insert $\Delta(L_0^i)$ into $L_0^i$} \label{code:l0:firstinsert}  \label{code:ins2}
		}\label{code:end:ilargerone} 
		
		\While{$i<k$ AND $\Delta(L_0^i)\neq \emptyset$}
		{
			\label{code:insert:while}
			READ($L_{i+1}^{|L_{i+1}|}$)  \textbf{// Read $\Omega(Q^{i+1})$ }\label{code:read:item}\\
			$\Delta(L_{0}^{i+1})=\Delta(L_0^i) \qjoin \Omega(L_{i+1}^{|L_{i+1}|})$ \label{code:further:join}\\
			INSERT($\Delta(L_0^{i+1})$, $L_0^{i+1}$) \textbf{// Insert $\Delta(L_{0}^{i+1})$ into $L_{0}^{i+1}$} \label{code:further:insert}  \label{code:ins3}\\ 
			$i++$ \label{code:iplusplus}
		}\label{code:b:further}
		\If{$\Delta(L_{0}^{k}) \neq \emptyset$}{
			\textbf{Report $\Delta(L_{0}^{k})$ as new matches of $Q$}
		}
	}
}
\end{algorithm}

\nop{
The following theorem states that the number of the partial matches recorded in the expansion list $L$ is minimum. 

\begin{theorem}\label{theorem:prefix:promising}
Given a TC-query $P$ with the timing sequence $\{\e_{1}$, $\e_{2}$,...,$\e_{k}\}$,namely, $\e_{j}$ $\prec \e_{{j+1}}$ $(1\leq j < k)$, no partial match recorded at the expansion list is a discardable partial match.
\end{theorem}
\myproof{
Assuming that the corresponding expansion list of $P$ is $L=$ $\{L^1$, $L^2$, ..., $L^k\}$, then for $1\leq i\leq k$, any partial match $g$ in $L^i$ is a match of $Preq(\e_{i})$ which means $g$ is not discardable.
}
}

\textbf{Case 2: Edge expiry.} When an edge $\sigma$ expires,  we can remove all expired partial matches (containing $\sigma$) in expansion list $L$ by scanning $L^{1}$ to $L^{j}$ where $L^{j}$ is the rightmost item in $L$ which contains expired partial matches\optionshow{}{(Lines \ref{code:b:tcdel}-\ref{code:e:tcdel} in Algorithm \ref{alg:del})}. 

\optionshow{}{

\begin{algorithm}[!h]\small \caption{DELETE($\sigma$)}
 \label{alg:del}
\KwIn{$\sigma$: an expired edge  to be deleted}
\KwIn{$L_i$ $=\{L_i^1$, $L_i^2$,...,$L_i^{|Q^i|}\}$: the expansion list for TC-subquery $Q^i$ ($1\leq i\leq k$)}
\KwIn{$L_0$ $=\{L_0^1$, $L_0^2$,...,$L_0^{|Q^i|}\}$: the expansion list over $\{Q^1$,$Q^2$,...,$Q^k\}$}
\KwOut{Adjusted partial matches after deleting $\sigma$}
\For{each $Q^i$ where $\sigma$ matches at least one query edge}
{ 
   
	\For{$j = $ $1$ to $|L_i|$} 
	{
		\label{code:b:tcdel}
	   \textbf{/* Delete partial matches containing $\sigma$ in $L_i^j$ */} \\
	   DELETE($\sigma$, $L_i^j$) \\
	   \If{no partial matches is deleted in $L_i^j$}{
		   BREAK  \label{code:e:tcdel}
	   }
	   
	} 
	If there are expired partial matches deleted from $L_i^{|L_i|}$, scan $L_0^i$ to $L_0^k$ to delete expired partial matches in $L_0$ (Similar to Lines \ref{code:b:tcdel}-\ref{code:e:tcdel}) \label{code:further:del}
}
\end{algorithm}

}

\subsection{Answering non-TC-queries} \label{sec:baseline:nontcq}
\begin{figure}[!h]
\newcommand{\mywidth}{0.5\linewidth}
\centering
	\resizebox{0.95\linewidth}{!}
	{
		\includegraphics{\picfolder 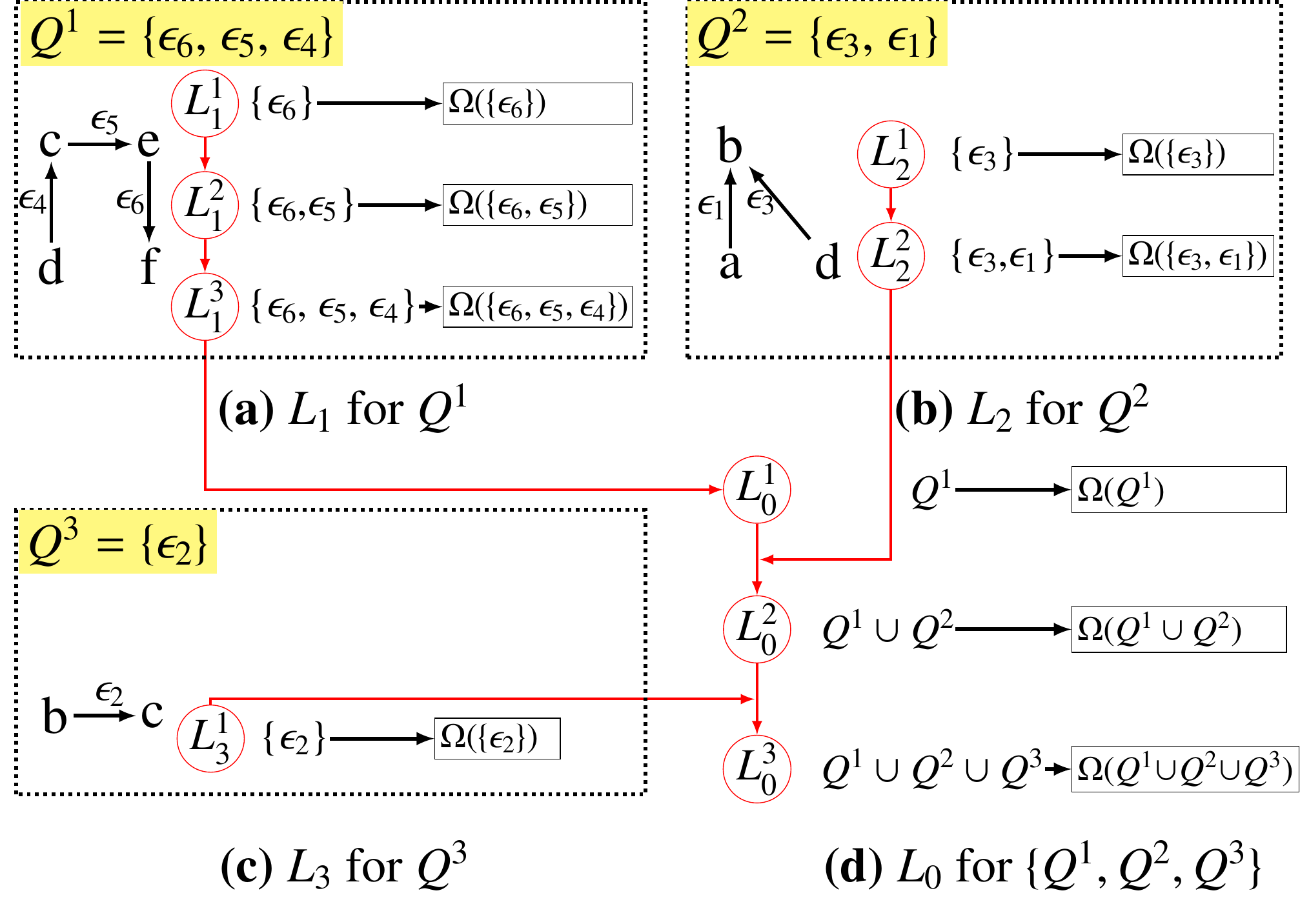}
	}

\caption{An TC decomposition of query $Q$}
\label{fig:telist:px}
\end{figure}

We decompose a non-TC-query $Q$ into a set of subqueries $D =\{Q^1$, $Q^2$,...$Q^k\}$, where each $Q^i$ is a TC-subquery, $Q=$ $\bigcup_{i=1}^{k}(Q^k)$ and there is no common query edge between any two TC-subqueries.  We call $D$ as a \emph{TC decomposition} of $Q$. The example query $Q$ is decomposed into $\{Q^1,Q^2,Q^3\}$, as shown in Figure \ref{fig:telist:px}. 
Since each TC-subquery $Q^i$ can be efficiently evaluated as described in the previous section, we focus on how to join those matches of $Q^i$ ($i=1,...,k$) into matches of $Q$ in the stream scenario.

For the sake of presentation, we assume that the decomposition of query $Q$ is given; decomposition is further discussed in Section \ref{sec:decompose:method}. We use $L_i$ $=\{L_1^1$, $L_i^2$,...,$L_i^{|E(Q^i)|}\}$ to denote the corresponding expansion list for each TC-subquery $Q^i$. Recall the definition of prefix-connected sequence  (Definition \ref{def:pre:conn:seq}).  
\mbred{We can find a permutation of $D$ whose prefix sequence always constitutes a weakly connected subquery of $Q$ as follows:  we first randomly extract a TC-subquery $Q^1$ from $D$; and then we extract a second TC-subquery $Q^2$ who have common vertex with $Q^1$ (Since $Q$ is weakly connected, we can always find such $Q^2$); repeatedly, we can always extract another TC-subquery from $D$ who have common vertex with some previously extracted TC-subquery and finally form a prefix-connected permutation of $D$. Without loss of generality, we assume that $\{Q^1$, $Q^2$,...,$Q^k\}$ is a prefix-connected permutation of $D$ where the subquery induced by \{$Q^1$, $Q^2$,..., $Q^i$\} is always weakly connected ($1\leq i\leq k$). }
Actually, the prefix-connected permutation corresponds to a join order, based on which, we can obtain $\Omega(Q)$ by joining matches of each $Q^i$. Different join orders lead to different intermediate result sizes, resulting in different performance.\optionshow{We do not discuss join order selection in this paper due to space constraints; this is a well-understood problem. We include our approach to the problem in the full paper \cite{fulltimingsubg}. For this paper, we assume that the prefix-connected sequence $D=\{Q^1$, $Q^2$,...,$Q^k\}$ is given. }{We discuss join order selection in Section \ref{sec:join:order}. Until then, we assume that the prefix-connected sequence $D=\{Q^1$, $Q^2$,...,$Q^k\}$ is given. }

\optionshow{}{
\begin{figure}[!h]
\newcommand{\mywidth}{0.6\linewidth}
\centering
	\resizebox{\mywidth}{!}
	{
		\includegraphics{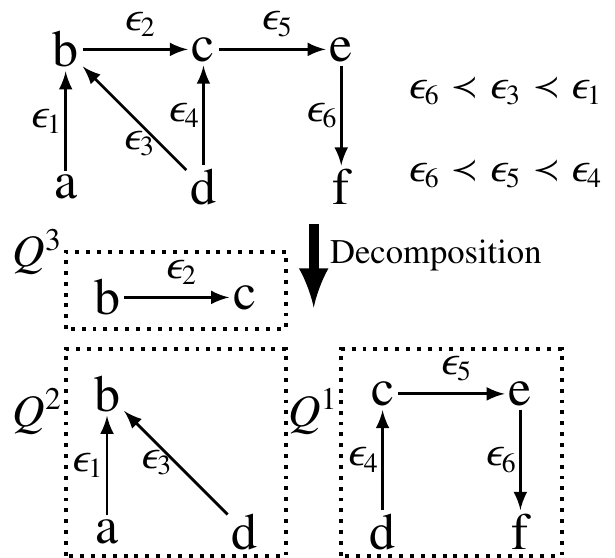}
	}
\caption{A TC decomposition of query $Q$}
\label{fig:tc:decompose}
\end{figure}
} 

For example, Figure \ref{fig:telist:px} illustrates a decomposition of query $Q$  ($Q^1$, $Q^2$, $Q^3$). We obtain the matches of $Q$ as $\Omega(Q)=\Omega(Q^1)\qjoin \Omega(Q^2)...\qjoin \Omega(Q^k)$. Like TC-query, we can also materialize some intermediate join results to speed up online processing. According to the prefix-connected sequence over $Q$, we can define the expansion list, denoted as $L_0$ for the entire query $Q$ (similar to TC-query). For example, the corresponding expansion list $L_0$ $=\{L_0^1$, $L_0^2$, $L_0^3 \}$ (for query $Q$) is given in Figure \ref{fig:telist:px}. Each item $L_0^i$ records the intermediate join results $\Omega(\bigcup\nolimits_{x = 1}^{i} {Q^x })$.

Assume that an incoming edge $\sigma$ contributes to new matches of TC-subquery $Q^i$ (denoted as $\Delta(L_i^{|L_i|})$) . If $i>1$, we let $\Delta(L_0^i)=\Delta(L_i^{|L_i|}) \qjoin \Omega(L_0^{i-1})$ (Line \ref{code:l0:firstjoin} in Algorithm \ref{alg:insert}). If $\Delta(L_0^i) \neq \emptyset$ , we insert $\Delta(L_0^i)$ into $L_0^i$ as new matches of $L_0^i$ . Then, $\Delta(L_0^i)$ $\qjoin \Omega(Q^{i+1})$ may not be empty and the join results (if any) are new partial matches that should be stored in $L_0^{i+1}$ ($\bigcup_{x=1}^{i+1}(Q^x)$). Thus, we need to further perform $\Delta(L_{0}^{i})$ $\qjoin \Omega(L_{i+1}^{|L_{i+1}|})$ to get new partial matches (denoted as $\Delta(L_{0}^{i+1})$) and insert them into  $L_0^{i+1}$ as new matches of $\bigcup_{x=1}^{i+1}(Q^x)$ . We repeat the above process until no new partial matches are created or the new partial matches are  exactly answers of the entire query $Q$ (Lines \ref{code:insert:while}-\ref{code:iplusplus}). Note that when partial matches of different subqueries are joined, we verify both  structure and timing order constraints.

When an edge $\sigma$ expires where $\sigma$ matches $\e \in$ $Q^i$, we discard all partial matches containing $\sigma$ in expansion list $L_i$ as illustrated previously. If there are expired matches for $Q^i$ (i.e., matches of $Q^i$ that contain $\sigma$), then we also scan $L_0^i$ to $L_0^k$ to delete partial matches containing $\sigma$.\optionshow{}{
(Pseudo codes for deletion are presented in Algorithm \ref{alg:del}).}

\section{Match-Store Tree} \label{sec:compression}
We propose a tree data structure, called match-store tree (MS-tree, for short), to reduce the space cost of storing partial matches in an expansion list.  Each tree corresponds to an expansion list.  Let's formally define MS-tree to present how the corresponding partial matches are stored and then illustrate how to access partial matches in MS-tree for the computation.

\begin{figure}[!h]
\newcommand{\mywidth}{0.60\linewidth}
\centering
	\resizebox{\mywidth}{!}
	{
		\includegraphics{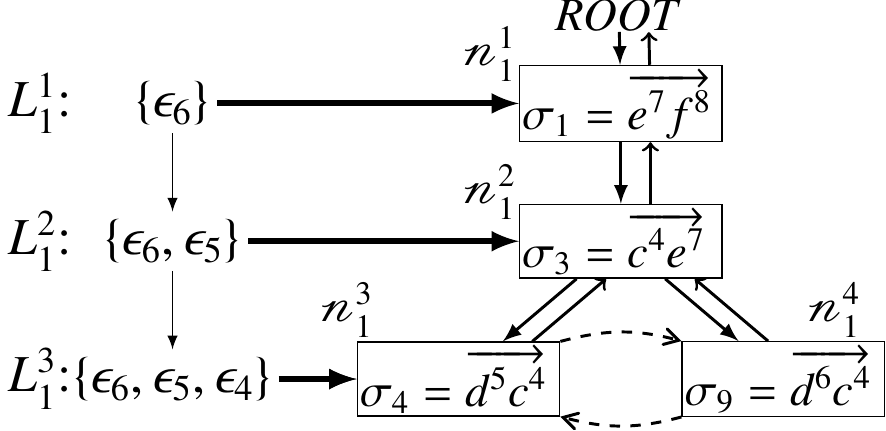}
	}
\caption{MS-tree of expansion list $L_1=$  $\{L_1^1$, $L_1^2$, $L_1^3\}$}
\label{fig:mstree}
\end{figure}

\vspace{-0.15in}

\subsection{Match-Store Tree} \label{sec:mstree:definition}
Consider an expansion list $L =$ \{$L^1$, $L^2$,...,$L^k$\} over timing sequence $\{\e_1$, $\e_2$,...,$\e_k\}$ where $L^i$ stores all partial matches of $\{\e_1$, $\e_2$,...,$\e_i \}$. For a match $g$ of $L^i$ ($1\leq i\leq k$), $g$ can be naturally presented in a \textbf{sequential form}: $\{\sigma_1$, $\sigma_2$,..,$\sigma_i\}$ where $g=$ $\bigcup_{j=1}^{i}(\sigma_j)$ and each $\sigma_{i'}$ ($1\leq i'\leq i$) is a match of $\e_{i'}$. Furthermore, $g^{\prime}$ $=g\setminus \{\sigma_i\}$ $=$ $\{\sigma_1$, $\sigma_2$,..,$\sigma_{i-1}\}$, as a match of $\{\e_1$, $\e_2$,...,$\e_{i-1} \}$, must be stored in $L^{i-1}$. Recursively, there must be $g^{\prime \prime}$ $=g^{\prime}\setminus \{\sigma_{i-1}\}$ in $L^{i-2}$. For example, see the expansion list in Figure \ref{fig:timing:expansion:list}. For partial match \{$\sigma_1$, $\sigma_3$, $\sigma_4$\} in item \{$\e_6$, $\e_5$, $\e_4$\}, there are matches \{$\sigma_1$, $\sigma_3$\} and \{$\sigma_1$\} in items \{$\e_6$, $\e_5$\} and \{$\e_6$\} of the expansion list, respectively. These three partial matches share a prefix sequence. Therefore, we propose a trie variant data structure to store the partial matches in the expansion list.

\begin{definition}[Match-Store Tree] \label{def:mstreedef}
Given a TC-query $Q$ with timing sequence $\{\e_1$,$\e_2$,...,$\e_k\}$ and the corresponding expansion list $L = $ $\{L^1$,$L^2$,...,$L^k\}$, the \emph{Match-Store tree} (\emph{MS-tree}) $M$ of $L$ is a trie variant 
built over all partial matches in $L$ that are in sequential form. Each node $\n$ of depth $i$ ($1\leq i\leq k$) in a MS-tree denotes a match of $\e_{i}$ and all nodes along the path from the root to node $\n$ together constitute a match of $\{\e_1$,$\e_2$,...,$\e_i\}$. Also, for each node $\n$ of a MS-tree, $\n$ records its parent node. Nodes of the same depth are linked together in a doubly linked list.
\end{definition}

For example, see the MS-tree for the expansion list for subquery $Q^1$ with the timing sequence \{$\e_6$, $\e_5$, $\e_4$\} in Figure \ref{fig:mstree}. 
The three matches (\{$\sigma_1$\} for node \{$\e_6$\}, \{$\sigma_1$, $\sigma_3$\} for node \{$\e_6$, $\e_5$\} 
and \{$\sigma_1$, $\sigma_3$, $\sigma_4$\} for node \{$\e_6$, $\e_5$, $\e_4$ \}) 
are stored only in a path ($\sigma_1$ $\rightarrow \sigma_3$ $\rightarrow \sigma_4$) in the MS-tree. Furthermore, partial match $\{\sigma_1$, $\sigma_3$, $\sigma_9 \}$ shares the same prefix path ($\sigma_1$ $\rightarrow \sigma_3$) with $\{\sigma_1$, $\sigma_3$, $\sigma_4\}$.
Thus, MS-tree greatly reduces the space cost for storing all matches by compressing the prefix.  
\optionshow{}{
Apparently, MS-tree can be seamlessly defined over the expansion list for the decomposition of a non-TC-query. For example, the MS-tree for expansion list $\{L_0^1$, $L_0^2$, $L_0^3\}$ for whole query $Q$ (see Figure \ref{fig:telist:px}) is shown in Figure \ref{fig:mszeronew}. For convenience, we use $M_i$ to denote the MS-tree for $L_i$ ($0\leq i \leq k$).

In fact, we can further reduce the space cost of the MS-tree $M_0$ for $L_0$. We know that each node in $M_0$ corresponding to a match of some TC-subquery. Consider a node $\n$ in $M_0$ that corresponds to a match $g$ of TC-subquery $Q_i$. $g$ has already been stored in MS-tree $M_i$ and we don't need to redundantly store $g$ in $\n$. Instead, we can just let $\n$  point to the leaf node corresponding to $g$ in $M_i$ and then we can easily access $g$ by backtracking the leaf node to the root in $M_i$. For example, the node $\n_0^1$ ($\n_0^2$) in Figure \ref{fig:mszeronew} can be easily replaced by a pointer pointing to leaf node $\n_1^3$ ($\n_1^4$) in Figure \ref{fig:mstree}.
} 


\optionshow{}{
\begin{figure}[!h]
\newcommand{\mywidth}{\linewidth}
\centering
	\resizebox{\mywidth}{!}
	{
		\includegraphics{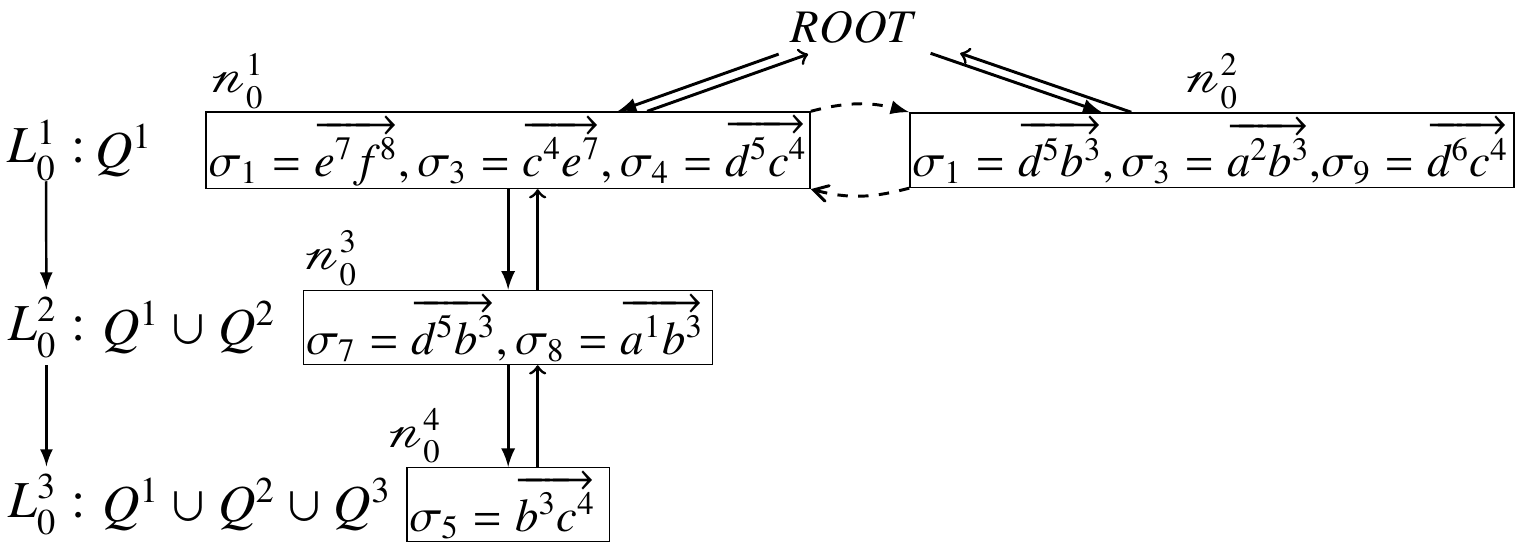}
	}
\caption{MS-tree of expansion list $L_0$ for \{$Q^1, Q^2, Q^3$\}}
\label{fig:mszeronew}
\end{figure}

} 

\subsection{MS-Tree Accessibility}\label{sec:mstree:access}
\optionshow{}{
Let's discuss how to access the partial matches stored in MS-tree. Note that these are basic serial accesses to MS-tree and the access issues in \emph{concurrent} mode will be discussed in Section \ref{sec:concurrency}. 
} 

Given an expansion list $L=$ \{$L^1$,$L^2$,...,$L^k$\} over timing sequence $\{\e_1$,$\e_2$,...,$\e_k\}$ and an MS-tree $M$ that stores all partial matches in $L$, there are three operations that $M$ needs to provide for computation: (1) reading all matches for some item $L^i$, i.e., $\Omega(L^i)$; (2) inserting a new match into some item $L^i$; (3) deleting expired partial matches (i.e.,partial matches containing expired edge). These three basic operations can be seamlessly applied to the MS-tree of expansion list $L_0$ over the decomposition of a non-TC-query.

{\textbf{Reading matches of $L^i$}:}
In a MS-tree, each $i$-length path starting from the root indicates a match of $L^i$, i.e., $\{\e_1$,$\e_2$,...,$\e_i \}$. We can obtain all matches of $L^i$ by enumerating all nodes of depth $i$ in $M$ with the corresponding doubly linked list, and then for each node of depth $i$, we can easily backtrack the $i$-length paths to get the match of $L^i$. Apparently, the time for reading partial matches in $L^i$ is $O(|L^i|)$ where $|L^i|$ denotes the number of partial matches in $L^i$.

\textbf{Inserting a new match of $L^i$:}
For a new match of $\{\e_1$,$\e_2$,...,$\e_i \}$: $g$ $=$ $\{\sigma_1$,  $\sigma_2$,...,$\sigma_i \}$ where each $\sigma_j$ matches $\e_j$, we need to insert a path $\{root\rightarrow\sigma_1$ $\rightarrow\sigma_2$...$\rightarrow\sigma_{i} \}$ into MS-tree. 
According to the insertion over expansion list,  $g$ must be obtained by $\{\sigma_1$,$\sigma_2$,...,$\sigma_{i-1} \}$ $\qjoin$  $\{\sigma_{i} \}$ and there must already be a path $\{root\rightarrow\sigma_1$ $\rightarrow\sigma_2$...$\rightarrow\sigma_{i-1} \}$ in MS-tree. Thus, we can just add $\sigma_i$ as a child of node $\sigma_{i-1}$ to finish inserting $g$.  For example, to insert a new match $\{\sigma_1$, $\sigma_3$, $\sigma_9\}$ of $\{\e_6$, $\e_5$, $\e_4 \}$, we only need to expand the path $\{root\rightarrow\sigma_1$ $\rightarrow\sigma_3\}$ by adding $\sigma_9$ as a child of $\sigma_3$ (see Figure \ref{fig:mstree}).
Note that, we can easily record node $\sigma_{i-1}$ when we find that $\{\sigma_1$,$\sigma_2$,...,$\sigma_{i-1} \}$ $\qjoin$  $\{\sigma_{i} \}$ is not $\emptyset$, thus inserting a match of $L^i$ cost $O(1)$ time. We can see that our insertion strategy does not need to wastefully access the whole path $\{root\rightarrow\sigma_1$ $\rightarrow\sigma_2$...$\rightarrow\sigma_{i-1} \}$ as the usual insertion of trie. 

\textbf{Deleting expired partial matches:} 
When an edge $\sigma$ expires, we need to delete all partial matches containing $\sigma$. Nodes corresponding to expired partial matches in MS-tree are called \emph{expired nodes} and we need to remove all expired nodes. Assuming that $\sigma$ matches $\e_i$, nodes containing $\sigma$ are exactly of depth $i$ in $M$. These nodes, together with all their descendants, are exactly the set of  expired nodes in $M$ according to the Definition of MS-tree. We first remove all expired nodes of depth $i$ (i.e., nodes which contain $\sigma$) from the corresponding doubly linked list, we further remove their children of depth $i+1$ from $M$. Recursively, we can remove all expired nodes from MS-tree. Consider the MS-tree in Figure \ref{fig:mstree}. When edge $\sigma_1$ (matching $\e_6$ in TC-query $\{\e_6,\e_5,\e_4\}$) expires, we delete node $\sigma_1$ in the first level of MS-tree, after which we further delete its descendant nodes $\sigma_3$, $\sigma_4$ and $\sigma_9$ successively. When an edge expired, the time cost for the deletion update is linear to the number of the corresponding expired partial matches.

\optionshow{
Although MS-tree is similar to trie, there are important differences between them. Due to space limits, we illustrate the difference in Section \ref{full-sec:mstree:vs:trie} of the full paper \cite{fulltimingsubg}.}
{
\subsection{MS-tree and Trie} \label{sec:mstree:vs:trie}
\emph{From the perspective of data structure:} Each node $\n$ in MS-tree, besides the links to $\n$'s children, there are extra links to $\n$'s father and siblings (doubly linked list). These extra links take an important role in reading matches of subqueries and avoiding inconsistency in the concurrent access over MS-tree (Section \ref{sec:concurrency}).

\emph{From the perspective of operation:} All operations (search/insertion/deletion) over trie always begin at the root, but we often access MS-tree horizontally. Each level of MS-tree is linked from the corresponding item in the expansion list. For example in Figure \ref{fig:mszeronew}, when reading $\Omega(Q^1 \cup Q^2)$, we begin accessing from $L_0^2$ (in the expansion list $L_0$) and obtain all matches $\Omega(Q^1 \cup Q^2)$ by enumerating all nodes at the 2-nd level in the MS-tree with the corresponding doubly linked list, and then for each such node, we can easily backtrack the paths to the root to obtain the match of $\Omega(Q^1\cup Q^2)$. 
} 


\section{Concurrency Management} \label{sec:concurrency}

To achieve high performance, the proposed algorithms can (and should) be executed in a multi-thread way. Since multiple threads access the common data structure (i.e., expansion lists) concurrently, there is a need for concurrency management. Concurrent computing over MS-tree is challenging since many different partial matches share the same branches (prefixes). We propose a fine-grained locking strategy to improve the throughput of our solution with consistency guarantee. We first introduce the locking strategy over the expansion list without MS-tree in Sections \ref{sec:lock:intuition} and \ref{sec:lock:solution} then illustrate how to apply the locking strategy over MS-tree in Section \ref{sec:lock:mstree}.


\subsection{Intuition}\label{sec:lock:intuition}
\begin{figure}[!h]
\newcommand{\mywidth}{0.2\linewidth}
\centering
	\resizebox{0.7\linewidth}{!}
	{
	    \includegraphics[scale=0.45]{\picfolder 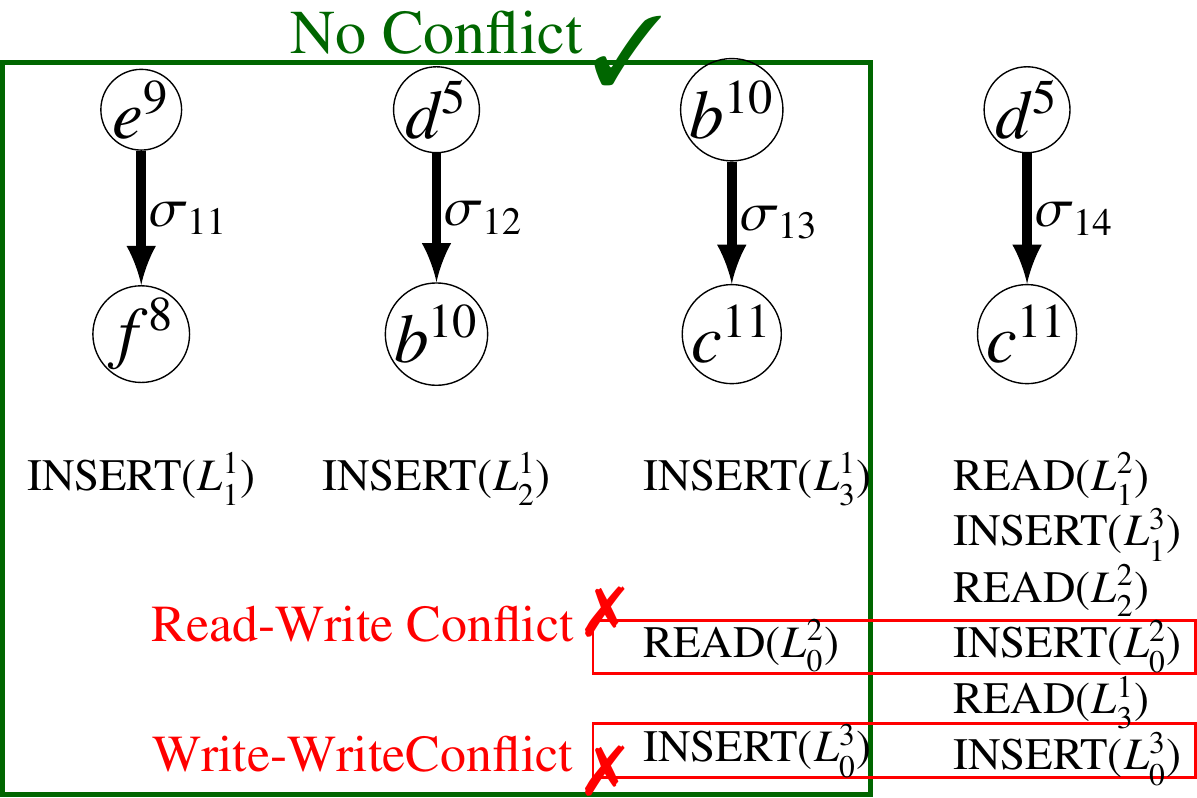}
	}	
\caption{Example of conflicts}
\label{fig:conflict:example}
\end{figure}

Consider the example query $Q$ in Figure \ref{fig:examplequery}, which is decomposed into three TC-subqueries $Q^1$, $Q^2$ and $Q^3$ (see Figure \ref{fig:telist:px}). Figure \ref{fig:telist:px} demonstrates expansion list $L_i$ of each TC-subquery $Q^i$ and the expansion list $L_0$ for the entire query $Q$. Assume that there are three incoming edges $\{\sigma_{11}, \sigma_{12}, \sigma_{13}\}$ (see Figure \ref{fig:conflict:example}) at consecutive time points. A conservative solution for inserting these three edges is to process each edge sequentially to avoid conflicts. However, as the following analysis shows, processing them in parallel does not lead to conflicts or wrong results. For convenience, insertion of an incoming edge $\sigma_i$ is denoted as $Ins(\sigma_i)$ while deletion of an expired edge $\sigma_j$ is denoted as $Del(\sigma_j)$.

Figure \ref{fig:conflict:example} illustrates the steps of handling each incoming edge based on the discussion in Section \ref{sec:baseline}. When $\sigma_{11}$ is inserted (denoted as $Ins(\sigma_{11})$), $\sigma_{11}$ matches query edge $\e_6$ and since $\e_6$ is the first edge in TC-subquery $Q^1$, we only need to insert match $\{\sigma_{11}\}$ into $\Omega(\e_{6})$ as the first item $L_1^1$ of expansion list $L_1$ (i.e., operation INSERT($L_1^1$)). Similarly, handling $Ins(\sigma_{12})$ where $\sigma_{12}$ matches $\e_3$ requires one operation:  INSERT($L_2^1$) (inserting $\{\sigma_{12}\}$ into $\Omega(\e_3)$). For $Ins(\sigma_{13})$ where $\sigma_{13}$ matches $\e_2$, we first insert $\sigma_{13}$ into $L_3^1$ (INSERT($L_3^1$)) as a new match of $Q^3$ (see Figure \ref{fig:telist:px}) and then we need to join $\{\sigma_{13}\}$ with $\Omega(Q^1\cup Q^2)$ (READ($L_0^2$)) and insert join results into $L_0^3$ (INSERT($L_0^3$)). Note that we consider the worst case in our analysis, namely, we always assume that the join result is not empty. Thus, to insert $\sigma_{13}$, we access the following expansion list items: INSERT($L_3^1$), READ($L_0^2$) and INSERT($L_0^3$).

Figure \ref{fig:conflict:example} shows that there is no common item to be accessed between $Ins(\sigma_{11})$, $Ins(\sigma_{12})$ and $Ins(\sigma_{13})$. Therefore, these incoming edges can be processed concurrently.

Let us consider an incoming edge $\sigma_{14}$ that matches $\{\e_4\}$, which is the last edge in the timing sequence of TC-subquery $Q^1$. According to Algorithm \ref{alg:insert}, we need to read $\Omega(\{\e_{6},\e_{5}\})$ and join $\Omega(\{\e_{6},\e_{5}\})$ with $\{\sigma_{14}\}$. Since $\e_4$ is the last edge in $Q^1$, if $\Omega(\{\e_{6},\e_{5}\})$  $\pjoin \{\sigma_{14}\}$ $\neq \emptyset$, the join results are new matches of $Q^1$, and will be inserted into $L_0^1$. As discussed in Section \ref{sec:baseline:nontcq}, we need to join these new matches of $Q^1$ with $\Omega(Q^2)$ resulting in new matches of $Q^1\cup Q^2$, which will be inserted into $L_0^2$. Finally, new matches of $Q^1\cup Q^2$ will be further joined with $\Omega(Q^3)$, after which new matches of $Q^1\cup Q^2\cup Q^3$ will be inserted into $L_0^3$. Thus, the series of operations to be conducted for $Ins(\sigma_{14})$ are as follows: READ($L_1^2$), INSERT($L_1^3$), READ($L_2^2$), INSERT($L_0^2$), READ($L_3^1$), INSERT($L_0^3$). Obviously, $Ins(\sigma_{14})$ may conflict with $Ins(\sigma_{13})$ since both of them will conduct INSERT($L_0^3$)  as indicated in Figure \ref{fig:conflict:example}. Thus, the concurrent execution requires a locking mechanism to guarantee the consistency.


\begin{definition}[Streaming Consistency]\label{def:stream:consistency}
\nop{ 
Consider a streaming graph $\iG$ with time window $W$, and a  query $Q$. The \emph{streaming consistency} requires the result of answering $Q$ at each time point to be the same as the result of executing operations (insertion/deletion) of edges in chronological order according to their timestamps.
} 
\mbred{Given a streaming graph $\iG$ with time window $W$ and a  query $Q$, the \emph{streaming consistency} requires that at each time point, answers of $Q$ are the same as the answers formed by executing insertion/deletion in chronological order  of edges.}
\end{definition}

Streaming consistency is different from \emph{serializability}, since the latter only requires the output of the concurrent execution to be equivalent to some serial order of transaction execution, while streaming consistency specifies that the order must follow the  timestamp order in $\iG$. For example, a concurrent execution that executes $Ins(\sigma_{14})$ followed by $Ins(\sigma_{13})$ would be serializable but would violate streaming consistency. 

\subsection{Locking Mechanism and Schedule}\label{sec:lock:solution}
We propose a locking mechanism to allow concurrent execution of the query execution algorithm while guaranteeing streaming consistency. The two main operations in streaming graphs, insertion of an incoming edge $\sigma$ (i.e., $Ins(\sigma)$) and deletion of an expired edge $\sigma^{\prime}$ (i.e., $Del(\sigma^{\prime})$), are modeled as \emph{transactions}. Each transaction has a timestamp that is exactly the time when the corresponding operation happens.  As discussed above, each edge insertion and deletion consists of elementary operations over items of the expansion lists, such as reading partial matches and inserting new partial matches. As analyzed in Section \ref{sec:lock:intuition}, concurrent execution of these operations may lead to conflicts that need to be guarded.

A naive solution is to lock all the expansion list items that may be accessed before launching the corresponding transaction. Obviously, this approach will degrade the system's degree of concurrency (DOC). For example, $Ins(\sigma_{13})$ and $Ins(\sigma_{14})$ conflict with each other only at items $L_3^1$, $L_0^2$ and $L_0^3$. The first three elementary operations of $Ins(\sigma_{13})$ and $Ins(\sigma_{14})$ can execute concurrently without causing any inconsistency. Therefore, a finer-granularity locking strategy is desirable that allows higher DOC while guaranteeing streaming consistency. For example, in Figure \ref{fig:conflict:example}, INSERT$(L_0^2)$ in $Ins(\sigma_{13})$ should be processed before the same operation in $Ins(\sigma_{14})$; otherwise, it will lead to inconsistency.


We execute each edge operation (inserting an incoming edge or deleting an expired edge) by an independent thread that is treated as a transaction, and there is a single main thread to launch each transaction. Items in expansion lists are regarded as ``resources'' over which threads conduct READ/INSERT/DELETE operations. Locks are associated with individual items in the expansion lists. An elementary operation (such as INSERT($L_3^1$) in $Ins(\sigma_{13})$) accesses an item  if and only if it has the corresponding lock over the item. The lock is released when the computation over $L^j$ is finished. Note that deadlocks do not occur since each transaction (thread) only locks at most one item (i.e., ``resource'') at a time.

\optionshow{}{
\label{sec:appendix:lockreq}
 \begin{figure}[!h]
\centering
\resizebox{\linewidth}{!}
{
	\includegraphics{\picfolder 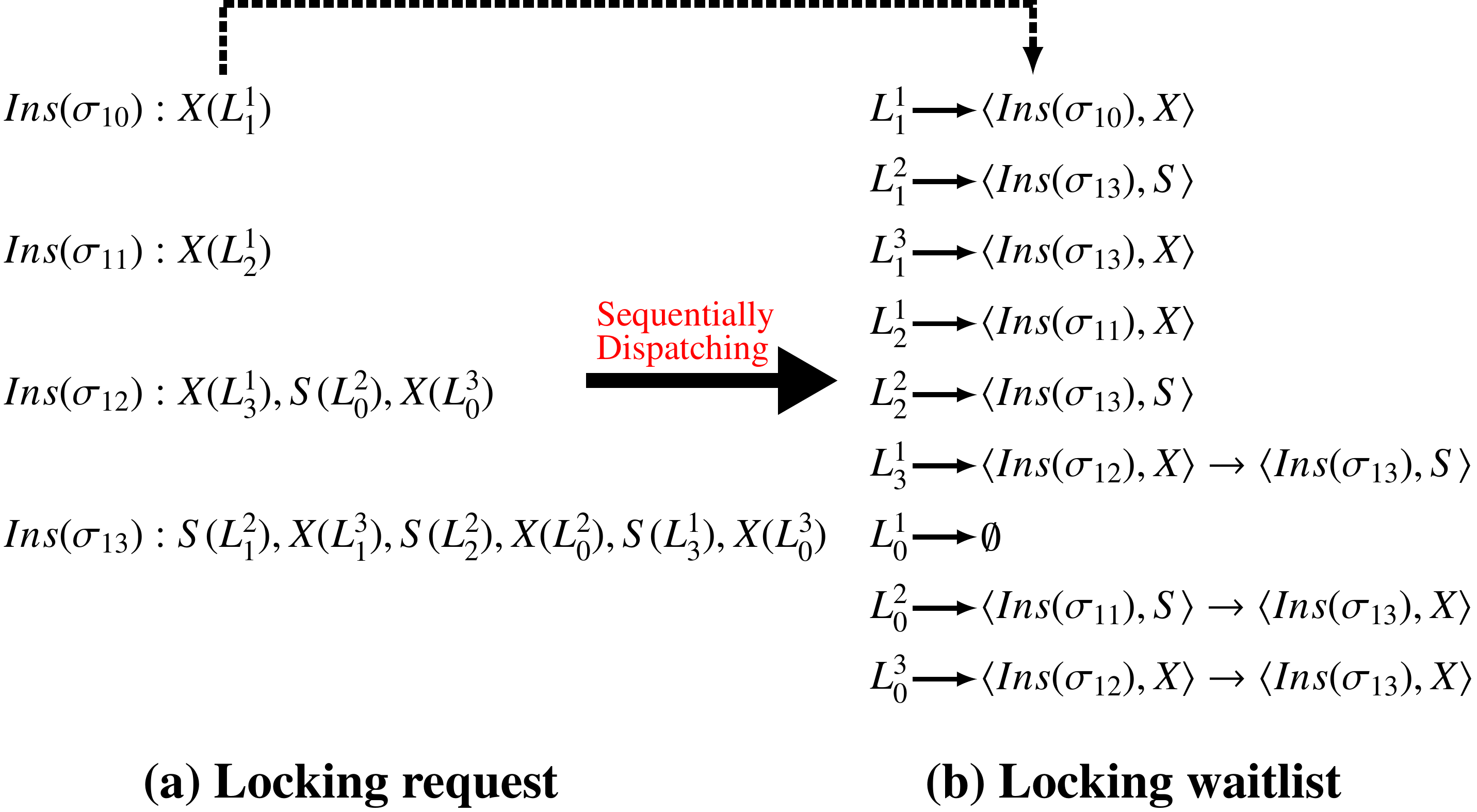}
}
\caption{Lock Request Dispatching}
\label{fig:intention:dispatch}
\end{figure}
}


\textbf{Main Thread.}
Main thread is responsible for launching threads\optionshow{.}{(see Algorithm \ref{alg:parallelmain}).}{}
Before launching a thread $T$, the main thread dispatches all \emph{lock requests} of $T$ to the \emph{lock wait-lists} of the corresponding items. Specifically, a lock request is a triple $\langle tID$, $locktype$, $L^j \rangle$ indicating that thread $tID$ requests a lock with type $locktype$ (shared -- $S$, exclusive -- $X$) over the corresponding item $L^j$ . For each item $L^j$ in expansion lists, we introduce a thread-safe wait-list consisting of all pending locks over $L^j$ sorted according to the timestamps of transactions in the chronological order.

\optionshow{}{
For example, we first extract all six lock requests in $Ins(\sigma_{12})$ in Figure \ref{fig:conflict:example} and dispatch them to the corresponding wait-list. }
Since there is a single main thread, the lock request dispatch as well as thread launch is conducted in a serial way. Hence, when a lock request of a thread is appended to wait-list of an item $L^j$, then those lock requests of previous threads for $L^j$ must have been in the wait-list since previous threads have been launched, which guarantees that lock requests in each wait-list are sorted in chronological order. Although thread launch is conducted in a serial way, once launched, all transaction threads are executed concurrently.  
\optionshow{}{
An example for lock request dispatching is presented in Figure \ref{fig:intention:dispatch}. 
}

\textbf{Transaction Thread execution.}
Concurrently processing insertion/deletion follows the same steps as the sequential counterparts except for applying (releasing) locks before (after) reading (READ) or writing (INSERT/DELETE)  expansion list items. Thus, in the remainder, we focus on discussing the lock and unlock processes. Note that, in this part, we assume that we materialize the partial matches ($\Omega(\cdot)$) using the naive representation (like Figure \ref{fig:timing:expansion:list}) without MS-tree. The locking strategy over MS-tree is more challenging that will be discussed in Sections \ref{sec:lock:mstree}.

Consider a thread $T$ that is going to access (READ/INSERT/ DELETE) an item $L^j$. $T$ can successfully obtain the corresponding lock of $L^j$ if and only if the following two conditions hold: (1)  the lock request of $T$ is currently at the head of the wait-list of $L^j$, and (2) the current lock status of $L^j$ is compatible with that of the request, namely, either $L^j$ is free or the lock over $L^j$ and the lock that $T$ applies are both shared locks.  Otherwise, thread $T$ will wait until it is woken up by the thread that just finishes computation on $L^j$. 

Once $T$ successfully locks item $L^j$, the corresponding lock request is immediately removed from the wait-list of $L^j$ and $T$ will conduct its computation over $L^j$. When the computation is finished, thread $T$ will release the lock and then wake up the thread (if any) whose lock request over $L^j$ is currently at the head of the wait-list.
 Finally, thread $T$ will continue its remaining computations.

\begin{theorem}\label{theorem:serializable}
The global schedule generated by the proposed locking mechanism is streaming consistent.
\end{theorem}
\optionshow{}{
\begin{proof}
Without loss of generality, consider two threads $T_1$ and $T_2$ which are launched at time $t_1$ and $t_2$, respectively ($t_1$ $< t_2$). Assume that $o_1$, $o_2$ are two consecutive operations of $T_1$ and $T_2$, respectively. If $o_1$ conflicts with $o_2$, $o_1$ will be executed before $o_2$ in the proposed scheme. Thus, if $o_1$ happened after $o_2$, then $o_1$ would not conflict with $o_2$ and we can swap the execution order of $o_1$ and $o_2$ without causing any inconsistency. Hence, the execution of $T_1$ and $T_2$ in the generated schedule can be always converted to the schedule where $T_2$ does not start until $T_1$ finishes without causing inconsistency. Consequently, the generated schedule is conflict-equivalent to the schedule where transactions are executed serially in chronological order based on their timestamps. Therefore, the transaction schedule obeys streaming consistency. 
\end{proof}
}

\optionshow{}{
\begin{algorithm}[h!]
\small
\caption{Parallel Processing Streaming Graphs}
 \label{alg:parallelmain}
\KwIn{Streaming graph $G$; Query Graph $Q$}
\KwOut{query results at each time point}
\For{each time point $t_i$}
{
$\sigma_i$ is an incoming edge at $t_i$;\\
$\sigma_j$ is an expired edge at $t_i$;\\
\If{$\sigma_i$ does not match any edge in query $Q$}
  {CONTINUE}
\Else 
   {Let $\Gamma$ be all lock requests for adding edge $\sigma_i$
   
   \For{each lock request in $\Gamma$}   
        {/*DISPATCH lock requests*/\\
        append it to the \emph{end} of the corresponding wait-list; }
    CREATE a new \textbf{thread} over Ins($\sigma_i$) (Algorithm \ref{alg:insert})
    }
 \If{$\sigma_j$ does not match any edge in query $Q$}
  {CONTINUE}
\Else 
   {Let $\Gamma$ be all lock requests for adding edge $\sigma_j$
   
    \For{each lock request in $\Gamma$}   
        {/*DISPATCH lock requests*/\\
        append it to the \emph{end} of the corresponding waiting list; }
          CREATE a new \textbf{thread} for Del($\sigma_j$) \optionshow{}{(Algorithm \ref{alg:del})}
   }
 }
\end{algorithm}


\begin{algorithm}[h!]
\small
\caption{Applies/releases S/X-lock}
 \label{alg:lock:process}
\KwIn{An item $L^i$ and the corresponding wait-list $waitlist(L^i)$}
\KwIn{Current thread $T$}
\KwOut{$T$ successfully applies/releases S/X-lock over $L^i$}
\DontPrintSemicolon 
\SetKwBlock{iFunc}{function}{end function}
\iFunc(apply\_S/X-lock{(}{)})
{
	\While{the lock request of $T$ is not at the head of $waitlist(L^i)$ OR the lock status of $L^i$ is exclusive}
	{
		$thread\_wait()$ \label{code:thread:wait}
	}\label{code:while:compat}
	apply S/X-lock over $L^i$ \\
	pop the head of $waitlist(L^i)$ 
}
\iFunc(release\_S/X-lock{(}{)})
{
	release S/X-lock over $L^i$ \label{code:release} \\
	If $waitlist(L^i)$ is not empty, wake up the thread whose lock request is at the head of $waitlist(L^i)$  \label{code:wakeup}
}
\end{algorithm}
} 

\subsection{Concurrent Access over MS-tree} \label{sec:lock:mstree}
Consider an expansion list $\{L^1$, $L^2$,...,$L^k\}$ whose partial matches are stored in MS-tree $M$. Each partial match of $L^i$ ($1\leq i\leq k$) exactly corresponds to a distinct node of depth $i$ in $M$. Thus, locking $L^i$ is equivalent to locking over all nodes of depth $i$ in $M$. Partial matches are not stored independently in MS-tree, which may cause inconsistency when concurrent accesses occur. For example, consider the MS-tree in Figure \ref{fig:mstree}. Assuming that a thread $T_1$ is reading partial matches of $\{\e_6$, $\e_5 \}$, $T_1$ will backtrack from node $\n_1^2$ (i.e., $\sigma_3$) to read  $\n_1^1$ (i.e., $\sigma_1$). Since $T_1$ only locks $L_1^2$, if another thread $T_2$ is deleting $\n_1^1$ at the same time, $T_2$ and $T_1$ will conflict. Therefore, we need to modify the deletion access strategy over the MS-tree to guarantee streaming consistency as follows.

Consider two threads $T_1$  and $T_2$ that are launched at time $t_1$ and time $t_2$ ($t_1$ $< t_2$), respectively. Assuming that $T_1$ is currently accessing partial matches of $L^{d_1}$ in $M$ while $T_2$ is accessing partial matches of $L^{d_2}$, let's discuss when inconsistency can happen. There are three types of accesses that each $T_i$ can perform and there are three cases for node depths $d_1$ and $d_2$ ($d_1<d_2, d_1=d_2$ and $d_1>d_2$). Thus, there are total $3\times3\times3$ $=27$ different cases to consider, but the following theorem tells us that only two of these cases will cause inconsistency in concurrent execution. 

\begin{theorem}\label{theorem:inconsistency}
Concurrent executions of $T_1$ and $T_2$ will violate streaming consistency if and only if one of these two cases occur:

\begin{enumerate}
\item 
	$d_1 > d_2$, $T_1$ reads partial matches of $L^{d_1}$ and $T_2$ deletes partial matches of $L^{d_2}$. When $T_1$ wants to read some node $\n$ during the backtrack to find the corresponding whole path, $T_2$ has already deleted $\n$, which causes the inconsistency.
\item
	$d_1>d_2$, $T_1$ inserts partial match $g=$ $\{\sigma_1$, $\sigma_2$,...,$\sigma_{d_1}\}$  of $L^{d_1}$ and $T_2$ deletes partial matches of $L^{d_2}$. When $T_1$ wants to add $\sigma_{d_1}$ as a child of $\sigma_{d_1-1}$, $T_2$ has deleted $\sigma_{d_1-1}$, which causes the inconsistency.
\end{enumerate}
\end{theorem}
\optionshow{}{
\begin{proof}
clearly, the two cases identified will cause inconsistency; let's discuss why the remaining $25$ cases will not cause inconsistency. When $d_1 = d_2$: $T_1$ and $T_2$ must be reading the same item and they will not conflict. Thus, we consider the cases where $d_1 \neq d_2$ or both $T_1$ and $T_2$ are reading partial matches.
\begin{enumerate}
\item
If $d_1 < d_2$:
	\begin{enumerate}
	\item
		If $T_1$ is reading partial matches of $L^{d_1}$: (1) if $T_2$ is inserting new partial matches into $L^{d_2}$, $T_2$ will only add new children for some nodes of depth $d_2-1$ and $T_1$ will not conflict with $T_2$; (2) if $T_2$ is deleting partial matches in $L^{d_2}$, the depth of expired nodes is not less than $d_2$ and $T_1$ will not conflict with $T_2$.
	\item
		If $T_1$ is inserting partial matches into $L^{d_1}$: if $T_1$ has lock request in the wait-list of $L^{d_2}$, then the lock request must be before that of $T_2$ ($T_1$ is launched before $T_2$), which means $T_2$ can not access $L^{d_2}$ until $T_1$ finishes accessing $L^{d_2}$ (in the future), thus, $T_1$ will not conflict with $T_2$. Otherwise, $T_1$ will not access partial matches in $L^{d_2}$ and (1) if $T_2$ is reading partial matches of $L^{d_2}$, the backtrack of $T_2$ will not involve the new nodes $T_1$ adds, and $T_1$ will not conflict with $T_2$; (2) if $T_2$ is not reading partial matches of $L^{d_2}$, there will be no common data that both $T_1$ and $T_2$ will access, thus, $T_1$ will not conflict with $T_2$.
	\item
		If $T_1$ is deleting partial matches in $L^{d_1}$, then $T_1$ must have lock request in the wait-list of $L^{d_2}$ before that of $T_2$ ($T_1$ is launched before $T_2$), and $T_2$ can not currently access $L^{d_2}$. Thus, $T_1$ will not conflict with $T_2$.
	\end{enumerate}
\item
If $d_2 < d_1$:
	\begin{enumerate}
	\item 
		If $T_2$ is reading partial matches of $L^{d_2}$, similar to the case where $d_1<d_2$, $T_1$ will not conflict with $T_2$.
	\item
		If $T_2$ is inserting partial matches into $L^{d_2}$, similar to the case where $d_1<d_2$, either $T_2$ can not access $L^{d_2}$ until $T_1$ finish accessing $L^{d_2}$ (in the future) or there is no common data that both $T_1$ and $T_2$ will access. Thus, $T_1$ will not conflict with $T_2$.
	\item
		If $T_2$ is deleting partial matches in $L^{d_2}$ and $T_1$ is deleting partial matches in $L^{d_1}$, then $T_1$ will not access the expired nodes $T_2$ delete and $T_1$ will not conflict with $T_2$.
	\end{enumerate}
\end{enumerate}
Thus, only two cases of the $27$ will cause inconsistency in our computation.
\end{proof}
} 

Theorem \ref{theorem:inconsistency} shows that inconsistency is always due to a thread $T_2$ deleting expired nodes that a previous thread $T_1$ wants to access without applying locks. However, if we make $T_2$ wait until previous thread $T_1$ finishes its execution, the degree of parallelism will certainly decrease. In fact, to avoid inconsistency, we only need to make sure that the expired nodes that $T_2$ wants to delete are invisible to threads launched later than $T_2$ while accessible to threads that are launched earlier. We achieve this by slightly modifying the deletion strategy over MS-tree with only negligible extra time cost. Specifically, consider the thread $T_2$ that deletes partial matches of $L^{d_2}$, when $T_2$ is going to delete expired node $\n_{d_2}$ of depth $d_2$ in $M$, $T_2$ does not ``totally'' remove $\n_{d_2}$ from $M$. Instead, $T_2$ ``partially'' removes $\n_{d_2}$ as follows: (1) $T_2$ removes $\n_{d_2}$ from the corresponding doubly linked list, and (2) $T_2$ disables the link (pointer) from $\n_{d_2}$'s parent to $\n_{d_2}$ while the link from $\n_{d_2}$ to its parent remains. 
\optionshow{}{
For example, Figure \ref{fig:partial:remove} demonstrates how an expired node $\n_{d_2}$ is partially removed. After $T_2$ partially removes all expired nodes, $T_2$ will finally remove all expired nodes from MS-tree $M$.

\begin{figure}[!h]
\newcommand{\mywidth}{0.9\linewidth}
\centering
	\resizebox{\mywidth}{!}
	{
		\includegraphics{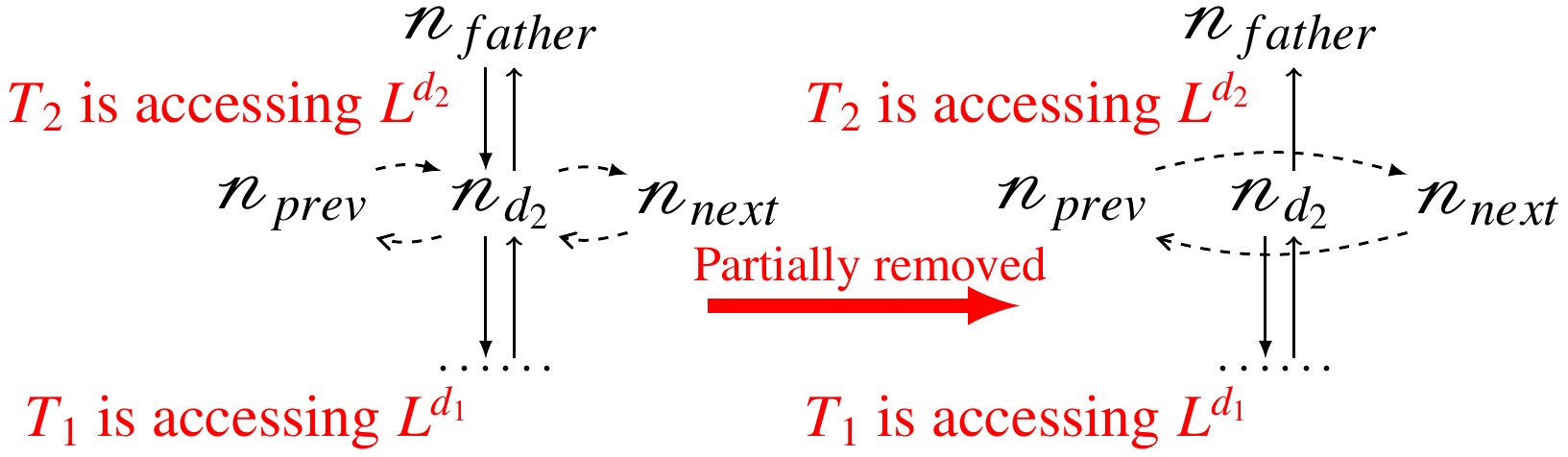}
	}
\caption{Demonstration of partially remove}
\label{fig:partial:remove}
\end{figure}
} 

\begin{theorem}\label{theorem:partial:remove}
Parallel accesses with modified deletion strategy over MS-tree do not result in streaming inconsistency. 
\end{theorem}
\optionshow{}{
\begin{proof}
Consider the two cases that will cause inconsistency (Theorem \ref{theorem:inconsistency}). For the first case, since $T_1$ is reading partial matches of $L^{d_1}$ ($d_1$ $>d_2$ and $t_1$ $<t_2$), $T_2$ does not partially remove expired nodes of depth $d_1$ yet, which means that $T_2$ does not totally remove any expired nodes according to the modified deletion strategy. Thus, $T_1$ can still safely access any nodes it wants to. When $T_2$ starts to totally remove expired nodes, it must has already partially removed all expired nodes -- that is, $T_2$ must have removed its lock requests from all $L^{d_i}$ where $d_i$ $>d_2$, which also means that threads launched before $T_2$ have already finished their computation over the MS-tree $M$ (Otherwise, $T_2$ would not have partially removed all expired nodes) and thus, it is safe for $T_2$ to finally remove all expired nodes.  Similarly, the second case will neither cause inconsistency.
\end{proof}
}

Our scheduling strategy over the MS-tree is different from the traditional tree protocol \cite{treelock1980silberschatz}. The classical tree protocol only guarantees the conflict equivalence to \emph{some} serial schedule, and there is no guarantee for \emph{streaming consistency} that requires a special serial order.

\section{Decomposition} \label{sec:decompose}
\optionshow{}{
For a non-TC-query $Q$, the query evaluation method in Section \ref{sec:baseline:nontcq} needs to decompose $Q$ into a set of TC-subqueries $D=\{Q^1,...,Q^k\}$. Although the query decomposition has been studied in subgraph query problem \cite{selectivityedbt15,sgi-messmer1998new}, none of these existing methods guarantees the decomposed subquery is a TC-query. Thus, none of them can be used in our problem.  
} 

We propose a \emph{cost model}-guided TC decomposition of query $Q$ based on the intuition that an incoming edge $\sigma$ should lead to as few join operations as possible. Cost of join operations varies in stream scenario and we only focus on the expected number of join operations to handle an incoming edge. Finding the most appropriate cost function is a major research issue in itself and outside the scope of this paper.

\subsection{Cost Model}
Assume that $Q$ has $|E(Q)|$ query edges $\e_j$ ($j$=1,...,$|E(Q)|$) and $Q$ is decomposed into $k$ TC-subqueries $Q^i$ ($i=1,...,k$). For simplicity, we assume that the probability of any incoming edge $\sigma$ matches each edge $\e_j$ in $Q$ is $1/d$, where $d$ is the number of distinct term edge labels (i.e., the label combining edge label and the connected node labels) in $Q$. Theorem \ref{theorem:join:number} tells us the expected number of join operation (in worst case) for an incoming edge.

\begin{theorem} \label{theorem:join:number}
Consider an incoming edge $\sigma$ that matches one or more edges in query $Q$. The total expected number of join operations for $Ins(\sigma)$ is
\[
N = \frac{1}{d}((|E(Q)| - 1) + \frac{k}{2}(k - 1))
\]
where $k$ is the number of TC-subqueries in the decomposition and $d$ is the number of distinct edge labels in $Q$.   
\end{theorem}
\optionshow{}{
\begin{proof}
Although an incoming edge may match more than one query edges in $Q$, the probability for each query edge  in $Q$ to be matched by an incoming edge is $1/d$  where $d$ is the number of distinct edge labels in $Q$.   
When an incoming edge $\sigma$ matches an edge $\e$ in $Q$, our method first computes the matches of TC-subquery $Q^i$ (Section \ref{sec:tcq:evaluation}) and then joins TC-subquery matches to find the matches of query $Q$ (Section \ref{sec:baseline:nontcq}). In the first step, if $\sigma$ matches the first edge $\e_1$ of the timing sequence $\{\e_1,...,\e_{|Q^j|}\}$ for a TC-subquery $Q^j$, there exists no join operation. If  $\sigma$ matches edge $\e_i$ ($i>1$), it needs to join with $\Omega(Preq(\e_{i-1}))$ that is recorded in MS-tree. Thus, it leads to one join operation. Assume that $Q$ is decomposed into $k$ TC-subqueries. Therefore, the expected number of join operations is
\[
N_1  = \frac{1}{d}(|E(Q)| - k){\rm{ = }}\frac{1}{d}(|E(Q)| - k)
\]
In the second step, we join TC-subquery matches to obtain matches of query $Q$. We consider the number of join operations \emph{in the worst case}. If  $\sigma$ contributes to a new match of TC-subquery $Q^1$, it needs to join all other TC-subqueries $Q^2$,...,$Q^k$. Therefore, it leads to $(k-1)$ join operations in the worst case. If $\sigma$ contributes to a new match of TC-subquery $Q^i$ ($i>1$), it needs to join $\Omega(\bigcup_{j=1}^{i-1}(Q^j))$, which is recorded in the upper levels of $M_0$ tree, and then join with the left TC-subqueries $Q^{i+1},...,Q^{k}$. Thus, if $i>1$, it leads to $(k-i)+1$ join operations. Therefore, the expected number of join operations in the second step is
\[
N_2  = \frac{1}{d}\sum\nolimits_{i = 2}^{k} {((k - i + 1)}  + (k - 1)) = \frac{1}{d}(\frac{{k^2  + k}}{2} - 1)
\]
Thus, the total expected number of join operations is
\[
N = N_1  + N_2  = \frac{1}{d}(|E(Q)| - 1 + \frac{k}{2}(k - 1))
\]
\end{proof}
}

Since $|E(Q)|$  and $d$ are fixed, the total expected number of join operations ($N$) increases with $k$. Therefore, we prefer to find a TC decomposition of size as small as possible.

\subsection{Decomposition Method}  \label{sec:decompose:method}
Given a query $Q$, to find a TC decomposition of size as small as possible, we propose the following solution.  We first extract all possible TC-subqueries of $Q$, denoted as $TCsub(Q)$. For a TC-subquery $Q^i$ of timing sequence $\{\e_1$,...,$\e_k\}$,  according to the definition of TC-query, any \emph{prefix} of the timing sequence constitutes a TC-subquery of $Q^j$. Thus, we can compute $TCsub(Q)$ by dynamic programming: 

\begin{enumerate}
\item
	We initialize $TCsub(Q)$ with all single edges of $Q$ since each single edge of $Q$ is certainly a TC-subquery of $Q$.
\item 
	With all TC-subqueries of $j$ edges, we can compute all TC-subqueries of $j+1$ edges as follows: for each TC-subquery $Q^i$ $=\{\e_1$,...,$\e_j\}$ with $j$ edges, we find all edges $\e_x$ such that $\e_j$ $\prec \e_x$. If $\e_x$ have common vertex with some $\e_{j^\prime}$ $(j^\prime \in [1, j])$, then we add $\{\e_1$,...,$\e_j$, $\e_x\}$ into $TCsub(Q)$ as a new TC-subquery of $j+1$ edges.

\item
	Repeat Step $2$ until there are no new TC-subqueries.
\end{enumerate}

\optionshow{}{The pseudocode is presented in Algorithm \ref{alg:tcsub}.} 
After computing $TCsub(Q)$, we need to compute a subset $D$ of $TCsub(Q)$ as a TC decomposition of $Q$, where the subset cardinality $|D|$ should be as small as possible. We use a greedy algorithm to retrieve the desired TC-subqueries from $TCsub(Q)$. We always choose the TC-subquery of maximum size from the remaining ones in $TCsub(Q)$ and there should be no common edges between the newly chosen subquery and those previously chosen ones.

\optionshow{}{
For example, consider query $Q$ in Figure \ref{fig:examplequery}. The corresponding $TCsub(Q)$ contains $10$ TC-subqueries: $\{\e_6, \e_5, \e_4\}$, $\{\e_3, \e_1\}$, $\{\e_5, \e_4\}$, $\{\e_6, \e_5\}$, $\{\e_1\}$, $\{\e_2\}$, $\{\e_3\}$, $\{\e_4\}$, $\{\e_5\}$, $\{\e_6\}$. 
We first choose $\{\e_6, \e_5, \e_4\}$ from $TCsub(Q_0)$, followed by $\{\e_3$, $\e_1\}$ and then $\{\e_2\}$. Thus, the TC decomposition of $Q$ will be $\{\{\e_6, \e_5, \e_4\}$, $\{\e_3$, $\e_1\}$, $\{\e_2\}$\} (Figure \ref{fig:telist:px}).
\optionshow{}{The pseudo code for conducting decomposition over $TCsub(Q)$ is presented in Algorithm \ref{alg:decomposition}.}
} 

\optionshow{}{
\begin{algorithm}[h!] \small \caption{Computing $TCsub(Q)$}
 \label{alg:tcsub}
\KwIn{Query $Q$: $\{\e_1$,...,$\e_m\}$}
\KwOut{The set of all TC-subqueries of $Q$: $TCsub(Q)$}
/* Initial a queue with all edges in $Q$ */ \\
Push each query edge $\e_i$ in $Q$ into $queue$ \\
\While{$queue$ is not empty}{
	$head\_subq = queue.pop()$ \\
	$TCsub(Q).add(head\_subq)$ \\
	\For{Each $\e_x$ $\in E(Q) - E(head\_subq)$}{
		\If{$\e_x$ is not adjacent to any edge in $head\_subq$}{
			CONTINUE
		}
		Let $\e_t$ be the last  item in $head\_subq$ \\
		\If{$\e_t \prec \e_x$}{
			Append $\e_x$ to $head\_subq$ \\
			$queue.push(head\_subq)$
		}
	}
}
RETURN \\
\end{algorithm}


\begin{algorithm}[!] \small \caption{Computing decomposition $D$ over $TCsub(Q)$}
 \label{alg:decomposition}
\KwIn{The set of TC-subqueries of $Q$: $TCsub(Q)$}
\KwOut{A decomposition of $Q$: $D$}
$D = \emptyset$ \\
Call $W=TCsub(Q)$\\
Sort  $W$ in ascending order of the number of query edges\\
\While{$D$ does not cover $Q$}{
	Find the TC-subquery $Q^i$ of most edges in $TCsub(Q)$ \\
	Remove $Q^i$ from $TCsub(Q)$ \\
	\If{$\exists Q^j\in D$ where $E(Q^i)\cap E(Q^j)\neq \emptyset$}{
		CONTINUE
	}
	Add $Q^i$ into $D$
}
RETURN \\
\end{algorithm}

} 

\optionshow{
}{

\subsection{Join order}\label{sec:join:order}
Given a decomposition $D=\{Q^1$,$Q^2$,...$Q^k\}$ of query $Q$, we obtain the matches of $Q$ as $\Omega(Q)=\Omega(Q^1)\Join \Omega(Q^2)...\Join \Omega(Q^k)$, in which the join order follows the prefix-connected sequence. Obviously, a good join order should lead to fewer intermediate results. Intuitively, we first find two TC-subqueries $Q^i$ and $Q^j$, where  $|\Omega(Q^i)\Join \Omega(Q^j)|$ is minimum among $D$ as the first two TC-subqueries in the targeted decomposition. Iteratively, we find other TC-subqueries resulting in small  intermediate result sizes. Join selectivity can be estimated according to the data distribution that has been  well-studied. However, this is infeasible for streaming graph data due to dynamic data distribution. Thus, we propose a simple yet effective heuristic rule.

\begin{definition}[Joint Number]\label{def:constraint:number}
Given two TC-subqueries $Q^i$ and $Q^j$ ($i\neq j$), the \emph{joint number} between $Q^i$ and $Q^j$, denoted as $JN(Q^i, Q^j)=n_v+n_t$, where $n_v$ is the number of common vertices between $V(Q^i)$ and $V(Q^j)$ and $n_t$ is the number of edge pairs $(\e_i$, $\e_j)$ 
 $\in E(Q^i) \times E(Q^j)$ such that there is timing order between $\e_i$ and $\e_j$.
\end{definition}

Given a decomposition $D=\{Q^1$,$Q^2$,...$Q^k\}$ of query $Q$, we first find two TC-subqueries $Q^1$ and $Q^2$ that are connected with each other, and the \emph{joint number} between them is maximum among all pairwise TC-subqueries. Iteratively, we find the TC-subquery $Q^3$, which is connected to $Q^1\cup Q^2$ and the \emph{joint number} between $Q^1\cup{Q^2}$ and $Q^3$ is maximum among all left TC-subqueries. We repeat the above process to find the whole  prefix-connected sequence over decomposition $D$ of query $Q$, which specifies the corresponding join oder. 
 

}

\section {Experimental Evaluation}
\label{sec:experiment}

We evaluate our solution against comparable approaches. All methods are implemented in C++ and run on a CentOS machine of 128G memory and two Intel(R) Xeon(R) E5-2640 2.6GHz CPU. Codes and query sets are available at \cite{timingsubggit}. We also present a case study in the full paper \cite{fulltimingsubg}.

\nop{
\begin{table}[!h]
\small
\centering
\caption{Summary of Datasets}     

\label{tab:dataset:stat}
  \begin{small}
    \resizebox{\textwidth}{!}
    {
    \begin{tabular}{|c|c|l|c|}
    \hline
{\bfseries \makecell{Datasets}} & {\bfseries \makecell{Type}} & {\bfseries \makecell{Vertexes}} & {\bfseries \makecell{Edges}} \\ \hline
Internet Backbone Traffic  & Network Flow & 2,601,005 & 445,440,480 \\ \hline
Linked Stream Benchmark  & Social Stream & 37,231,144 & 209,549,677 \\ \hline
    \end{tabular}
    }
\end{small} 
\end{table}
}

\subsection{Datasets}
\label{sec:setup}

\optionshow
{
We use three datasets  in our experiments: real-world network traffic dataset, wiki-talk network dataset and synthetic social stream benchmark. Due to space limits, we only report the experimental results over network dataset and social stream in this paper and that of wiki-talk are presented in the full paper \cite{fulltimingsubg}.
\textbf{The network traffic data} is the ``CAIDA Internet Anonymized Traces 2015 Dataset'' obtained from \url{www.caida.org}, which contains 445,440,480 communication records (edges) concerning 2,601,005 different IP addresses (vertices). 
\textbf{The wiki-talk dataset} is from the Standford SNAP library \cite{snapdataset} where a directed edge indicates that a user edit another user's talk page at a certain time point. This dataset contains 1,140,149 vertices and 7,833,140 edges. 
\textbf{Linked Stream Benchmark} \cite{lsbenchcode} is a synthetic streaming social graph data on user's traces and posts information. This dataset contains 209,549,677 edges and 37,231,144 vertices.  
}
{ 
We use three datasets  in our experiments: real-world network traffic dataset, wiki-talk network dataset and synthetic social stream benchmark. 

\textbf{The network traffic data} is the ``CAIDA Internet Anonymized Traces 2015 Dataset'' obtained from \url{www.caida.org}. The network data contains 445,440,480 communication records (edges) concerning 2,601,005 different IP addresses (vertices). Each edge is associated with a timestamp indicating the communication time. 

A network communication record is a five-tuple that includes the source IP address/port number, the destination IP address/port number and the protocol in use. We transform these five-tuples into a vertex/edge labelled streaming graph. Note that although we only study the vertex-labelled graphs in this paper, it is straightforward to extend our method to edge-labelled graphs. The vertex label is fixed as ``IP''. Each edge label is a triple $\langle$source port, destination port, protocol$\rangle$. Since the source ports vary a lot resulting in very low matching ratio of query edges, we  replace source port by a wildcard ``*'' that can match any source port. In fact, there are 65520 different destination ports where the top $6$ (i.e., the top $0.01\%$) frequent ports exist in more than $50\%$ (i.e., 222,720,240) communication records.  

\textbf{The wiki-talk dataset} is from the Standford SNAP library \cite{snapdataset} where a directed edge (A, B, $t$) indicates that user A edit user B's talk page at time $t$. This dataset contains 1,140,149 vertices and 7,833,140 edges and the total time span is 2,320 days. We use the first character of the user's name to be the label of a vertex.

\textbf{Linked Stream Benchmark} \cite{lsbench2012linked} is a synthetic streaming social graph data. There are three different components in the streaming data. The \emph{GPS stream} contains user's trace specified by longitudes, latitudes and the corresponding tracking time. The \emph{Post stream} contain posts from users and the information of photos uploaded by users. We set the user number parameter as 1 million and the entire time span of the streaming data as 10 days. Parameters except the user number and time span are applied in the default setting.  The streaming social data contains 209,549,677 edges and 37,231,144 vertices. The data generator is available in Google Code \cite{lsbenchcode}.  Each record in the streaming data is a five-tuple consisting of subject type/id, predicate, object type/id. We also build a streaming graph over the streaming data where vertex labels are the corresponding subject/object types and edge labels are the predicates. 
} 

\subsection{Query Generation}
\label{sec:querygen}
\optionshow
{
We generate query graphs by random walk over the data graph. For each subgraph $g$ that is retrieved from data graph, we need to further generate the timing order. In fact, there is a full timing order between any two edges in $g$  according to their inherent timestamps in the data graph. Hence, we can generate a subset of this full timing order to be that of $g$. We create a random permutation of $g$'s edges and then for any two edges $\e_i$, $\e_j \in E(g)$, we set $\e_i$ $\prec \e_j$ if and only if (1) $\e_i$ is before $\e_j$ in the permutation and (2)  the timestamp of $\e_i$ in $g$ is less than that of $\e_j$. The average selectivities of these queries are reported in 
Figure \ref{full-fig:selectivity} of the full paper \cite{fulltimingsubg}.
}
{
A usual method for generating query graph is to perform the random walk over the data graph. However, generating query graph with timing order constraints is non-trivial since we need to make sure that (1) timing order should be generated inherently with randomness to be representative and (2) query graph with the timing order should still have embedding (including the chronological order between edges) in the data graph. We propose a method generating queries satisfying these conditions and the corresponding average selectivities of these queries are reported in 
Figure \ref{fig:selectivity} in Section \ref{sec:appendix:selectivity}.
When generating a query,  we first conduct a random walk over the data graph and retrieve a subgraph $g$ where we generate random permutation of edges, assumed as \{$\e_1$, $\e_2$, ..., $\e_k$\}, then we set $\e_i \prec$ $\e_j$ if and only if (1) $\e_i$ is before $\e_j$ in the permutation; and (2) the timestamp of $\e_i$ in $g$ is less than that of $\e_j$.
In this way, we generate a query $Q$ with graph structure $g$ and timing order $\prec$. For edges in $g$, the positional order in random permutation and the order that they appear in the stream (chronological order of the corresponding timestamps) are independent of each other and hence the way we create timing order $\prec$ is of randomness. In this way, the query $Q$ we generate not only guarantees the representativeness but also makes sure that there exist subgraph in data graph that satisfies both time order and structure constraints of $Q$.
}

We generate $300$ queries over each dataset in our experiments. For each dataset, we set six different query sizes: $6$, $9$, $12$, $15$, $18$, $21$. For each query size, we generate $10$ query graphs by random walks over data graph. For each query graph $g$, we create $5$ different timing orders over $g$ where one is set as full order, one is set as $\emptyset$ and the other three are created by random permutations as illustrated previously.  


\subsection{Comparative Evaluation}
\label{sec:exp:com:eval}
Since none of the existing works support concurrent execution, all codes (including ours) are run as a single thread; the evaluation of concurrency management is in Section \ref{sec:exp:concurrency}. 
Our method, denoted as {Timing}, is compared with a number of related works. {SJ-tree} \cite{selectivityedbt15} is the closest work to ours. Since it does not handle the timing order constraints, we  verify answers from SJ-tree posteriorly with the timing order constraints. IncMat \cite{fan2013incremental} conducts static subgraph isomorphism algorithm when update happens over streaming graph.  We apply three different state-of-the-art static subgraph isomorphism algorithms to IncMat, including {\quicksi} \cite{quicksi2008},  {\turboiso} \cite{turboiso}, {\boostiso} \cite{boostiso}. \mbred{These methods are conducted over the affected area (see \cite{fan2013incremental}) window by window.} To evaluate the effectiveness of MS-tree, we also compare our approach with a counterpart without MS-trees (called {Timing-IND}) where every partial match is stored independently. 

There are $5$ different window sizes in our experiments: $10K$, $20K$, $30K$, $40K$ and $50K$ where each unit of the window size is the average time span between two consecutive arrivals of data edges in the dataset (i.e., the ratio of the total time span of whole dataset to the total number of data edges).

We evaluate the systems by varying window size $|W|$  and query size $|E(Q)|$. 
\optionshow
{In Section \ref{full-sec:exp:sizek} of the full paper \cite{fulltimingsubg}}{In Section \ref{sec:exp:sizek}}, 
we also compare our methods with comparative ones when varying the decomposition size $k$. The reported throughput (The number of edges handled per second) and space under a given group settings are obtained by averaging those from the corresponding generated queries.

\subsubsection{Time Efficiency Comparison}
\label{sec:exp:time}

\begin{figure}[h!]
\centering
    \begin{subfigure}[t]{0.45\linewidth}
	    \centering
	    \resizebox{\linewidth}{!}
	    {
	        \includegraphics{\timefolder 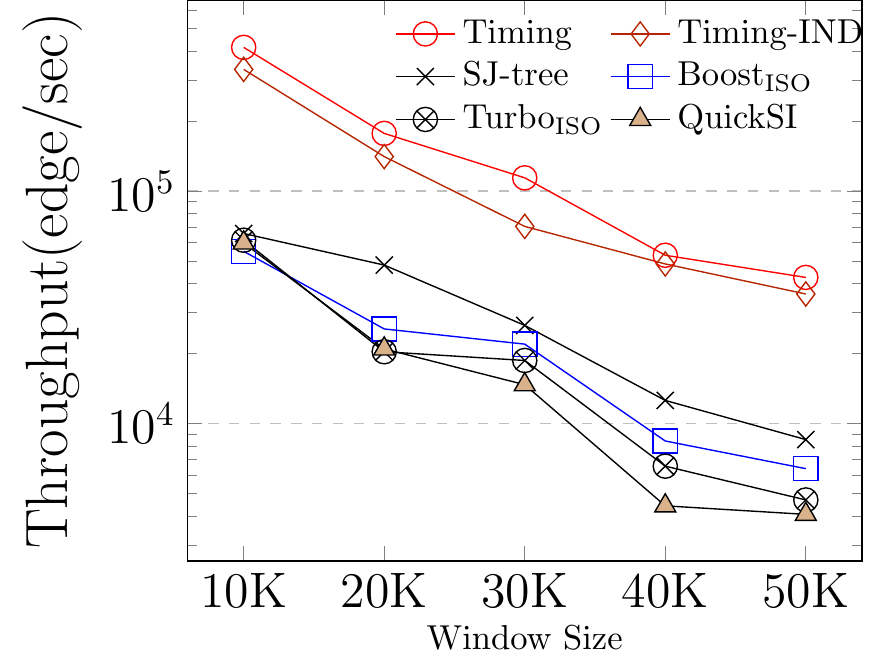}
	    }
	    \caption{Network Flow}
	    \label{fig:time:win:net}
    \end{subfigure}
      	\hspace{0.1in} 
    \begin{subfigure}[t]{0.45\linewidth}
        \centering
        \resizebox{\linewidth}{!}
        {
            \includegraphics{\timefolder 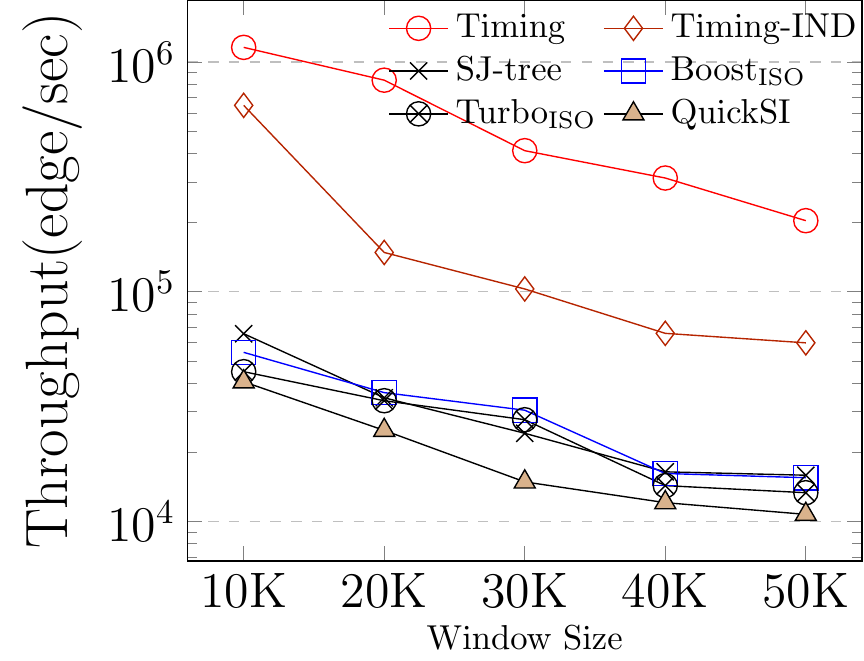}
        }
        \caption{Social Stream}
        \label{fig:time:win:rdf}
    \end{subfigure}
\optionshow{}{
     \begin{subfigure}[t]{0.45\linewidth}
         \centering
         \resizebox{\linewidth}{!}
         {
             \includegraphics{\timefolder 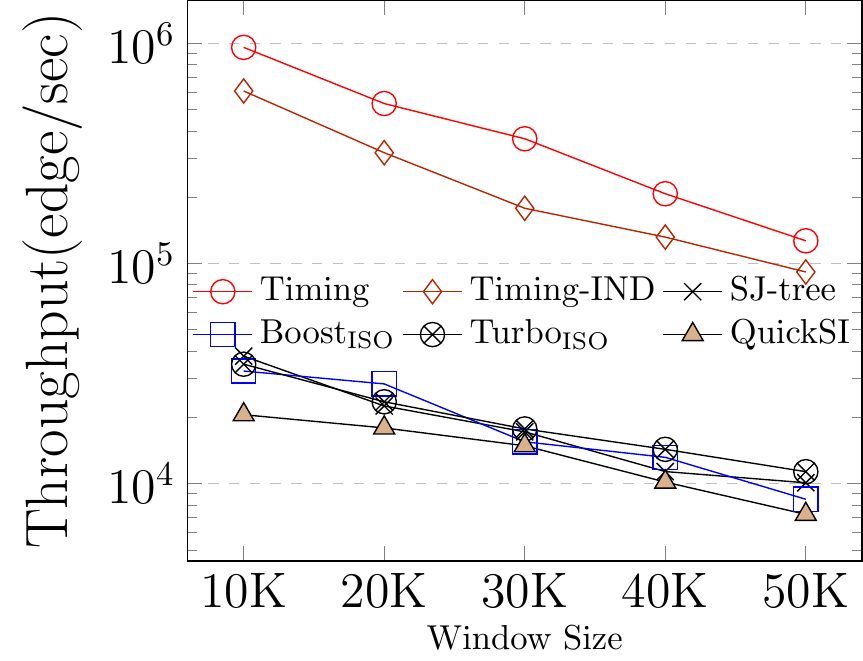}
         }
         \caption{Wiki-talk}
         \label{fig:time:win:wkt}
     \end{subfigure}
     }
\caption{Throughput over Different Window Size}
\label{fig:time:win}
\end{figure}
\vspace{-0.15in}
\begin{figure}[h!]
\centering
    \begin{subfigure}[t]{0.45\linewidth}
	    \centering
	    \resizebox{\linewidth}{!}
	    {
	        \includegraphics{\timefolder 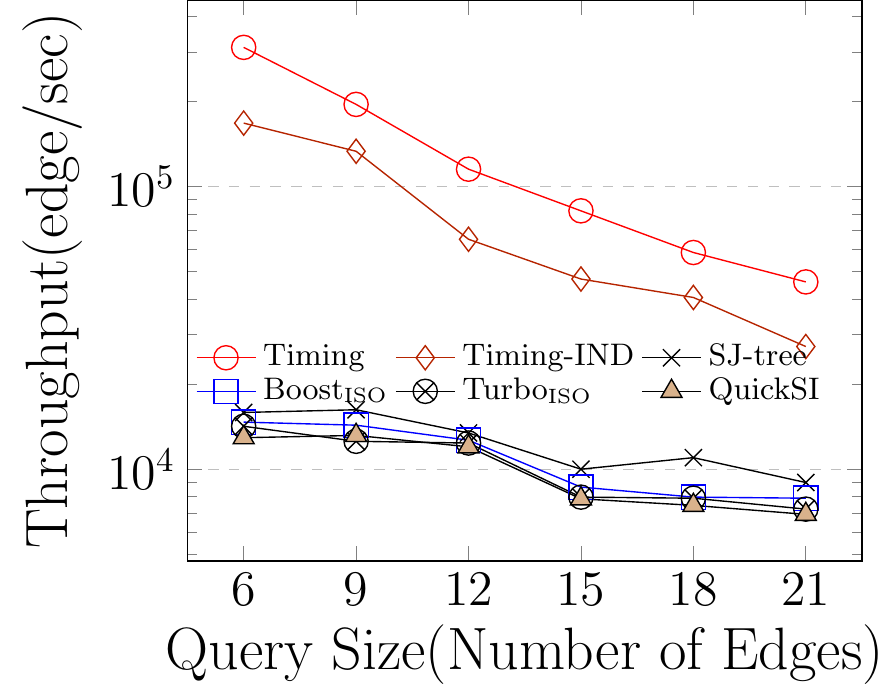}
	    }
	    \caption{Network Flow}
	    \label{fig:time:qsz:net}
    \end{subfigure}
      	\hspace{0.1in} 
    \begin{subfigure}[t]{0.45\linewidth}
        \centering
        \resizebox{\linewidth}{!}
        {
            \includegraphics{\timefolder 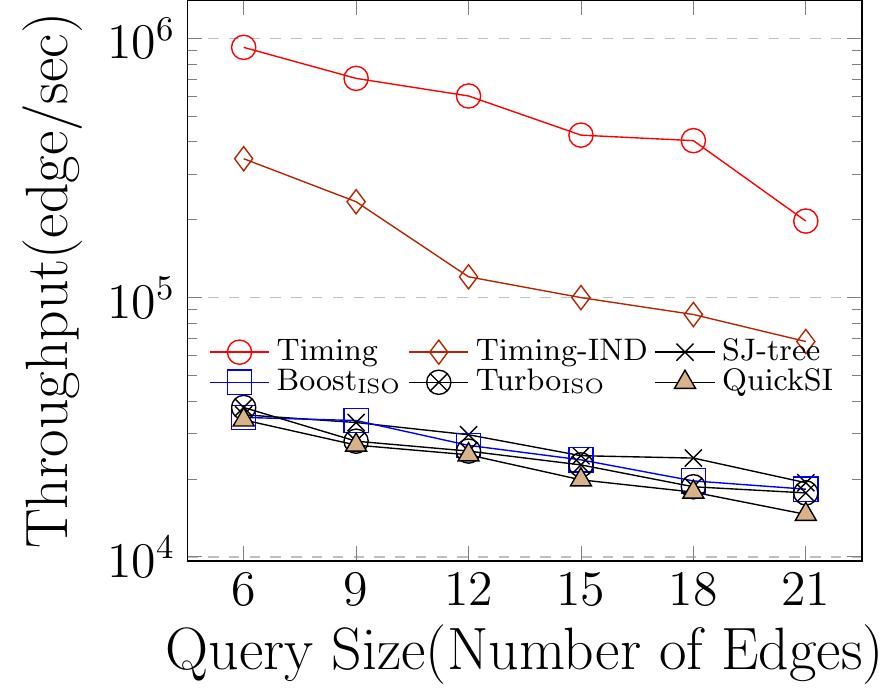}
        }
        \caption{Social Stream}
        \label{fig:time:qsz:rdf}
    \end{subfigure}
     \optionshow{}{
     \begin{subfigure}[t]{0.45\linewidth}
         \centering
         \resizebox{\linewidth}{!}
         {
             \includegraphics{\timefolder 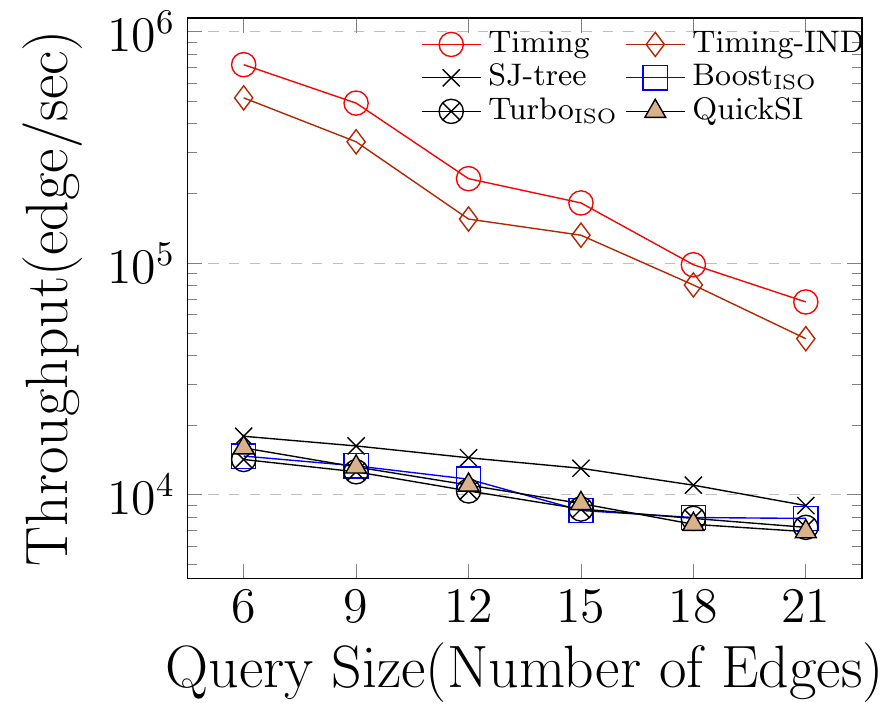}
         }
         \caption{Wiki-talk}
         \label{fig:time:qsz:wkt}
     \end{subfigure}
     } 
\caption{Throughput over Different Query Size}
\label{fig:time:qsz}
\end{figure}
\vspace{-0.1in}

Figures \ref{fig:time:win}-\ref{fig:time:qsz}  show that our method is clearly faster than other approaches over different window sizes and query sizes, respectively.  The reason for the superior performance of our method lies in two aspects. First, our method can filter out lots of discardable partial matches based on the timing order constraint. Second is the efficiency of MS-tree maintenance algorithms. For example, the deletion algorithm is linear to the total number of expired partial matches; while in SJ-tree, all partial matches need to be enumerated to find the expired ones. SJ-tree needs to maintain lots of discardable partial matches that can be filtered out by our approach. Furthermore, SJ-tree needs post-processing for the timing order constraint, which also increases running time. 
Finally, since Timing-IND does not use MS-tree to optimize the space and maintenance cost, it is not as good as Timing, as shown in our experiments. 

\begin{figure}[h!]
\centering
    \begin{subfigure}[t]{0.45\linewidth}
	    \centering
	    \resizebox{\linewidth}{!}
	    {
	        \includegraphics{\spacefolder 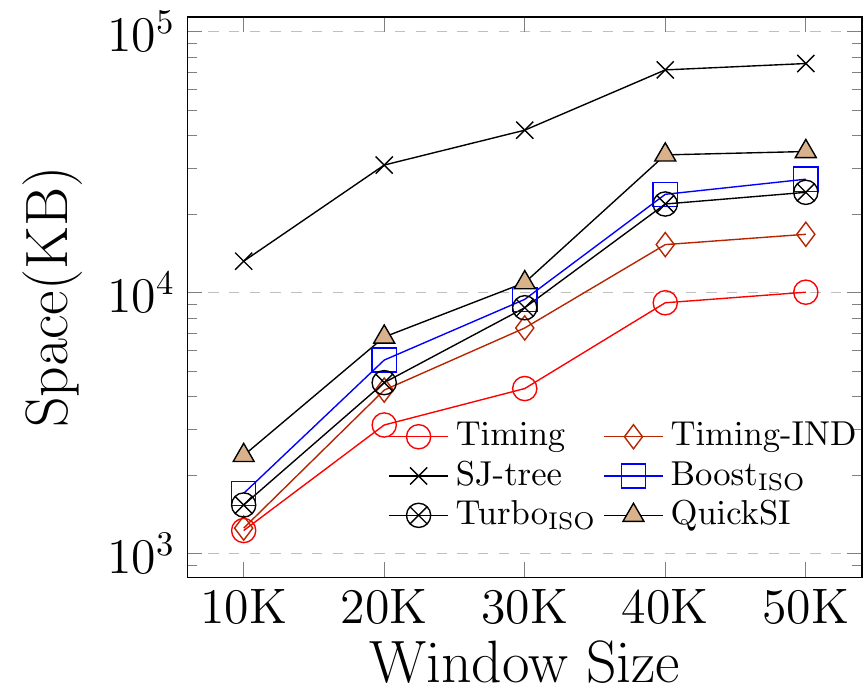}
	    }
	    \caption{Network Flow}
	    \label{fig:space:win:net}
    \end{subfigure}
      	\hspace{0.1in} 
    \begin{subfigure}[t]{0.45\linewidth}
        \centering
        \resizebox{\linewidth}{!}
        {
            \includegraphics{\spacefolder 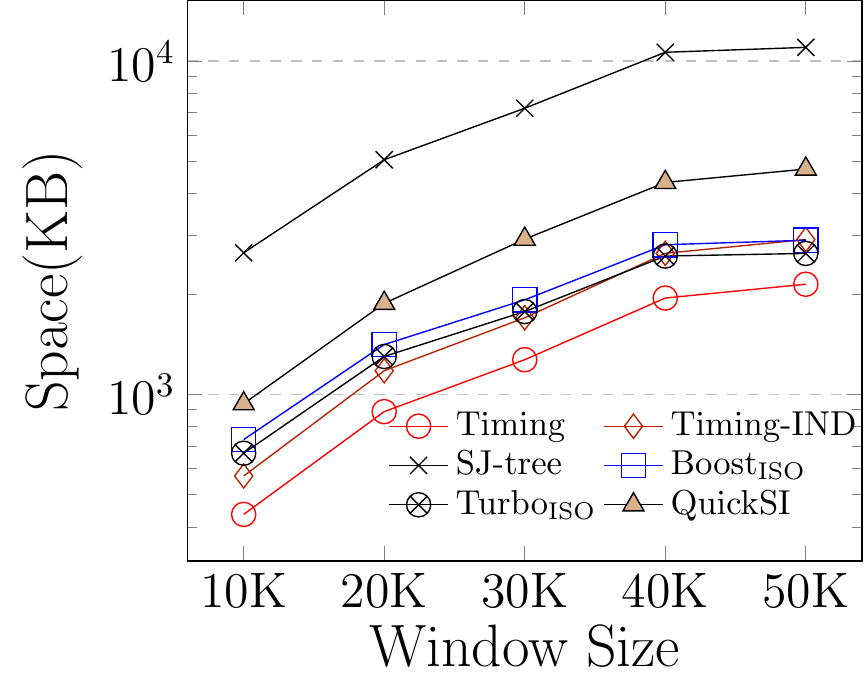}
        }
        \caption{Social Stream}
        \label{fig:space:win:rdf}
    \end{subfigure}
     \optionshow{}{
     \begin{subfigure}[t]{0.45\linewidth}
         \centering
         \resizebox{\linewidth}{!}
         {
             \includegraphics{\spacefolder 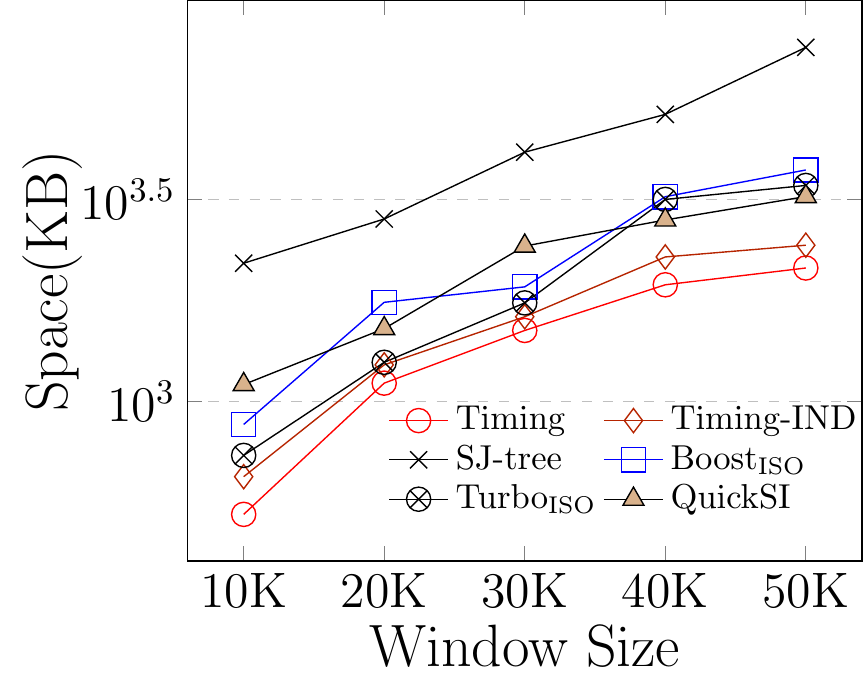}
         }
         \caption{Wiki-talk}
         \label{fig:space:win:wkt}
     \end{subfigure}
     }
\caption{Space over Different Window Size}
\label{fig:space:win}
\end{figure}
\vspace{-0.2in}
\begin{figure}[h!]
\centering
    \begin{subfigure}[t]{0.45\linewidth}
	    \centering
	    \resizebox{\linewidth}{!}
	    {
	        \includegraphics{\spacefolder 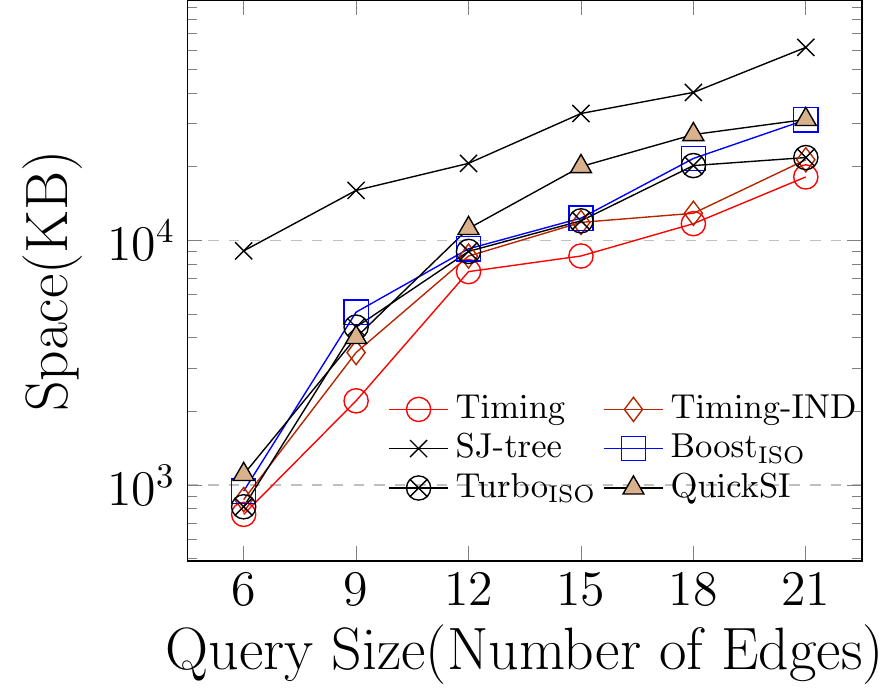}
	    }
	    \caption{Network Flow}
	    \label{fig:space:qsz:net}
    \end{subfigure}
      	\hspace{0.1in} 
    \begin{subfigure}[t]{0.45\linewidth}
        \centering
        \resizebox{\linewidth}{!}
        {
            \includegraphics{\spacefolder 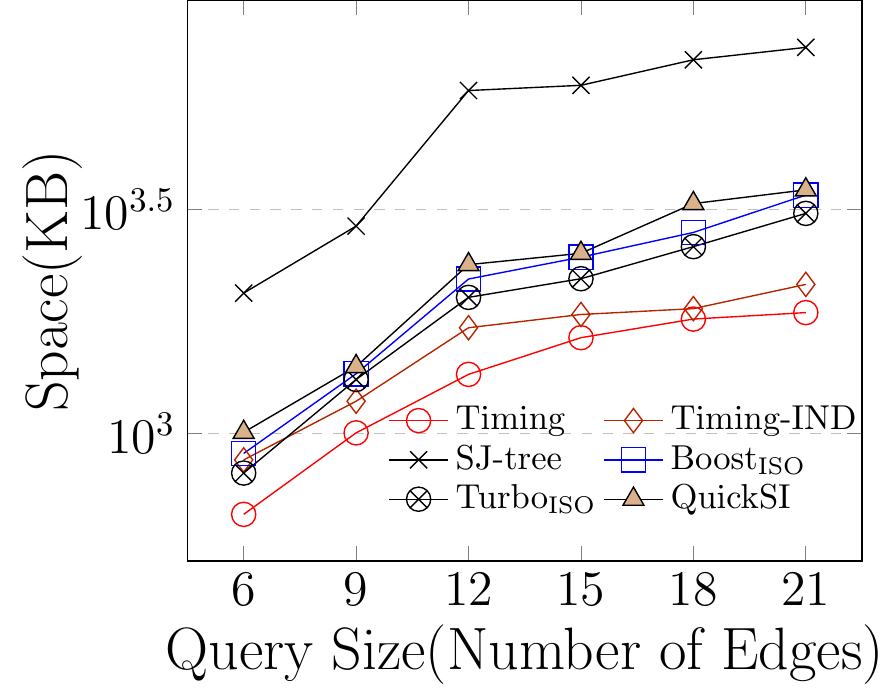}
        }
        \caption{Social Stream}
        \label{fig:space:qsz:rdf}
    \end{subfigure}
     \optionshow{}{
     \begin{subfigure}[t]{0.45\linewidth}
         \centering
         \resizebox{\linewidth}{!}
         {
             \includegraphics{\spacefolder 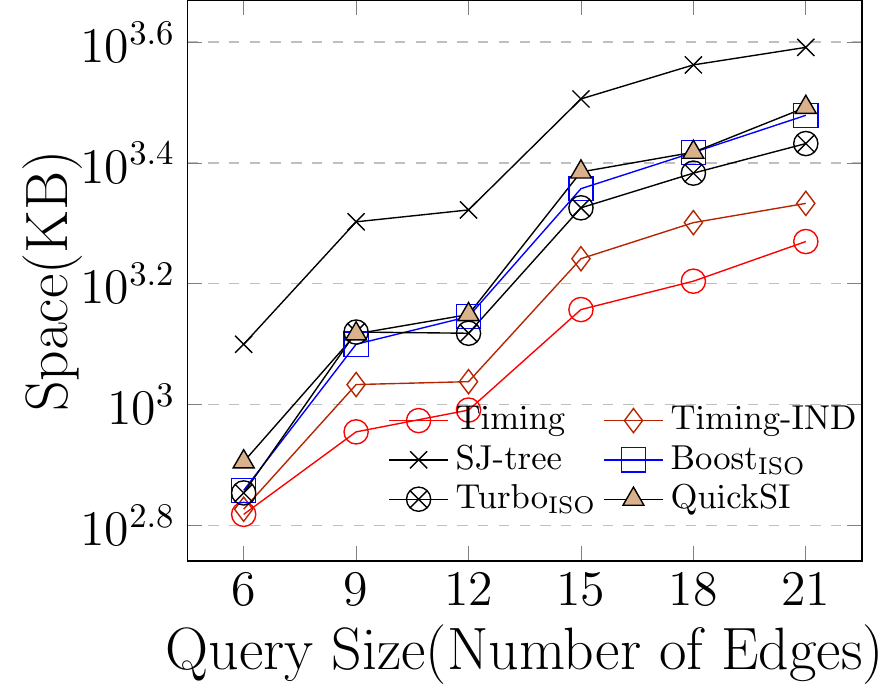}
         }
         \caption{Wiki-talk}
         \label{fig:space:qsz:wkt}
     \end{subfigure}
     }
\caption{Space over Different Query Size}
\vspace{-0.05in} 
\label{fig:space:qsz}
\end{figure}


\subsubsection{Space Efficiency Comparison}
\label{sec:exp:space}
We compare the systems with respect to their space costs. Since the streaming data in the time window changes dynamically, we use the average space cost in each time window as the metric of comparison, as shown in Figures \ref{fig:space:win}-\ref{fig:space:qsz}. We can see that both Timing-IND and Timing have much lower space cost than comparative approaches.  Our method is more efficient on space than SJ-tree because SJ-tree does not reduce the discardable partial matches, which wastes space. Our method only maintains partial matches without graph structure in the time window. However, \quicksi{}, \turboiso{} and \boostiso{} need to maintain the graph structure (adjacent list) in each window to conduct search. Also, these comparative methods can not reduce discardable edges that will never exist in any partial match, which results in wasting space.

\subsection{Concurrency Evaluation}
\label{sec:exp:concurrency}
\begin{figure}[h!]
\centering
    \begin{subfigure}[t]{0.43\linewidth}
	    \centering
	    \resizebox{\linewidth}{!}
	    {
	        \includegraphics{\concurfolder 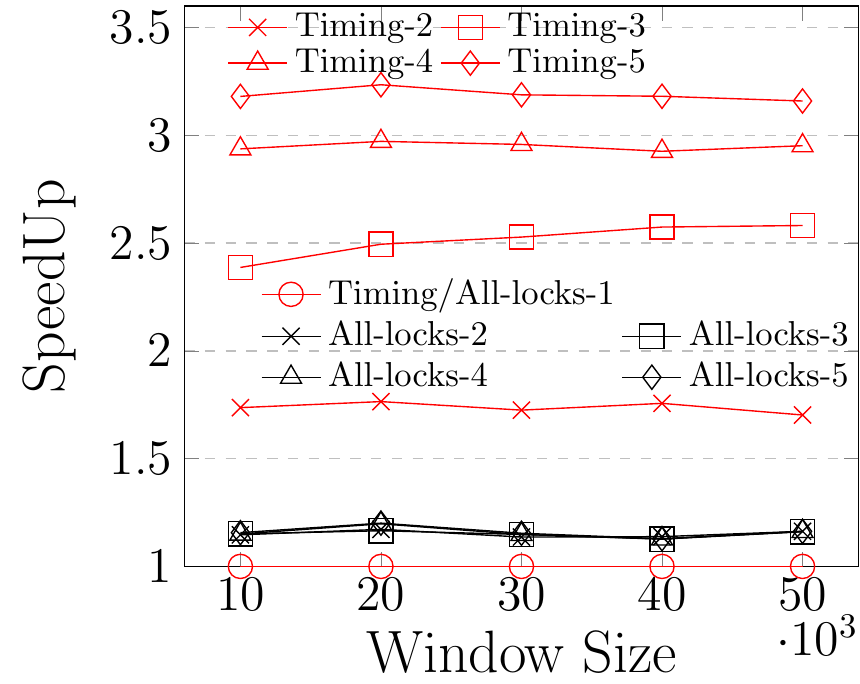}
	    }
	    \caption{Network Flow}
	    \label{fig:spd:win:net}
    \end{subfigure}
      	\hspace{0.1in} 
    \begin{subfigure}[t]{0.43\linewidth}
        \centering
        \resizebox{\linewidth}{!}
        {
            \includegraphics{\concurfolder 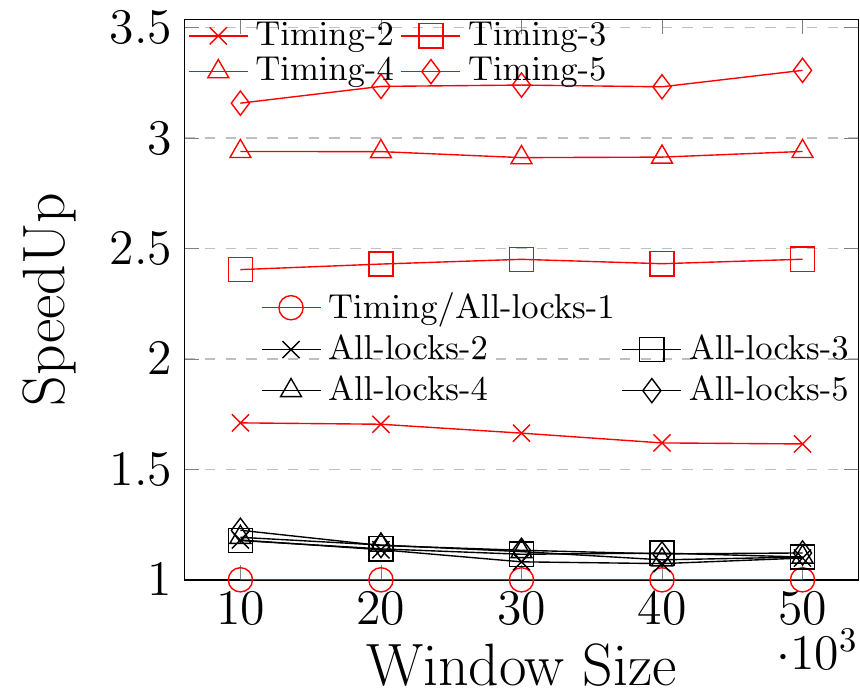}
        }
        \caption{Social Stream}
        \label{fig:spd:win:rdf}
    \end{subfigure}
     \optionshow{}{
     \begin{subfigure}[t]{0.43\linewidth}
         \centering
         \resizebox{\linewidth}{!}
         {
             \includegraphics{\concurfolder 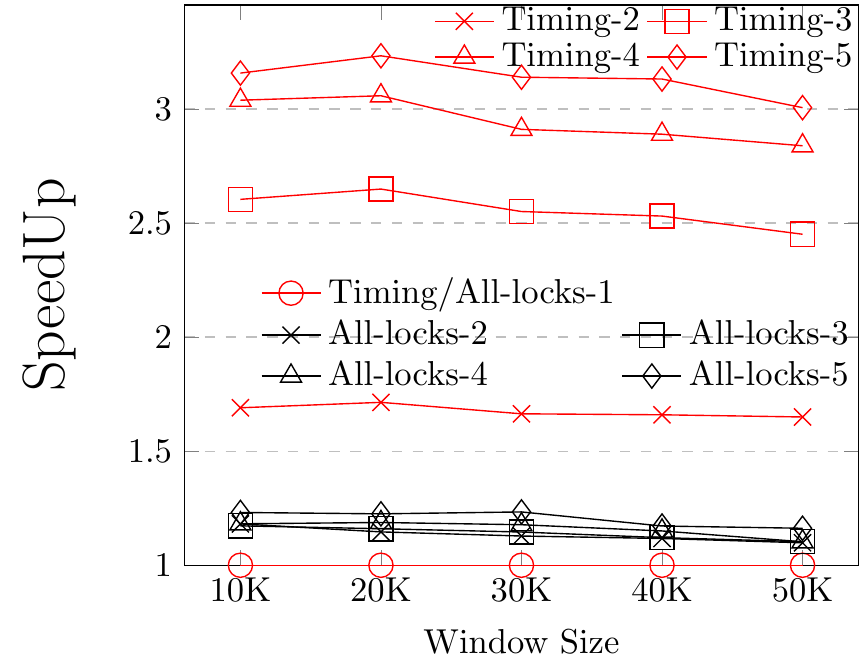}
         }
         \caption{Wiki-talk}
         \label{fig:spd:win:wkt}
     \end{subfigure}
     }
\caption{Speedup over Different Window Size}
\vspace{-0.15in}
\label{fig:spd:win}
\end{figure}

\begin{figure}[h!]
\centering
    \begin{subfigure}[t]{0.43\linewidth}
	    \centering
	    \resizebox{\linewidth}{!}
	    {
	        \includegraphics{\concurfolder 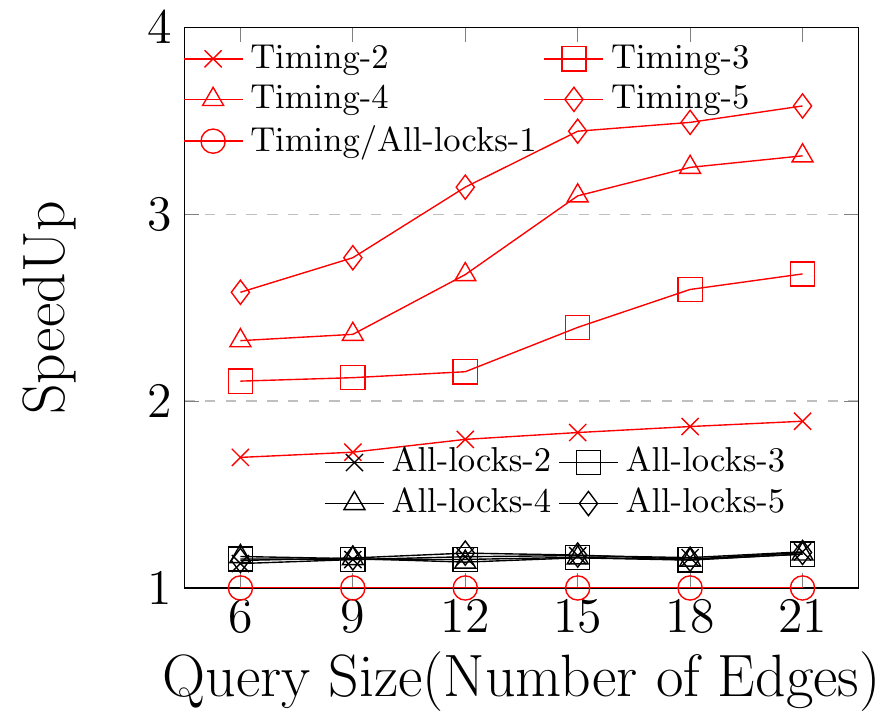}
	    }
	    \caption{Network Flow}
	    \label{fig:spd:qsz:net}
    \end{subfigure}
      	\hspace{0.1in} 
    \begin{subfigure}[t]{0.43\linewidth}
        \centering
        \resizebox{\linewidth}{!}
        {
            \includegraphics{\concurfolder 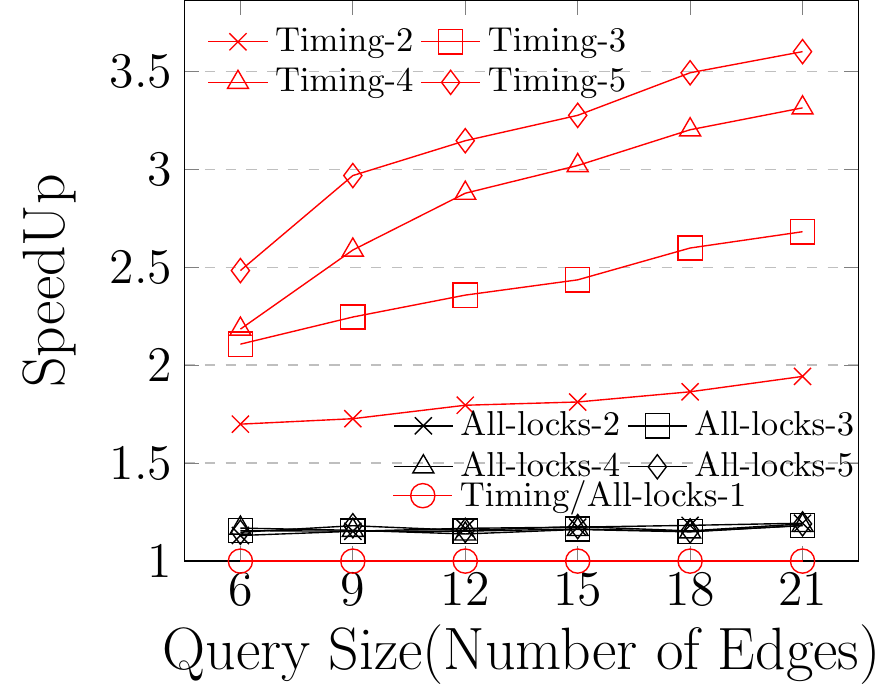}
        }
        \caption{Social Stream}
        \label{fig:spd:qsz:rdf}
    \end{subfigure}
     \optionshow{}{
     \begin{subfigure}[t]{0.43\linewidth}
         \centering
         \resizebox{\linewidth}{!}
         {
             \includegraphics{\concurfolder 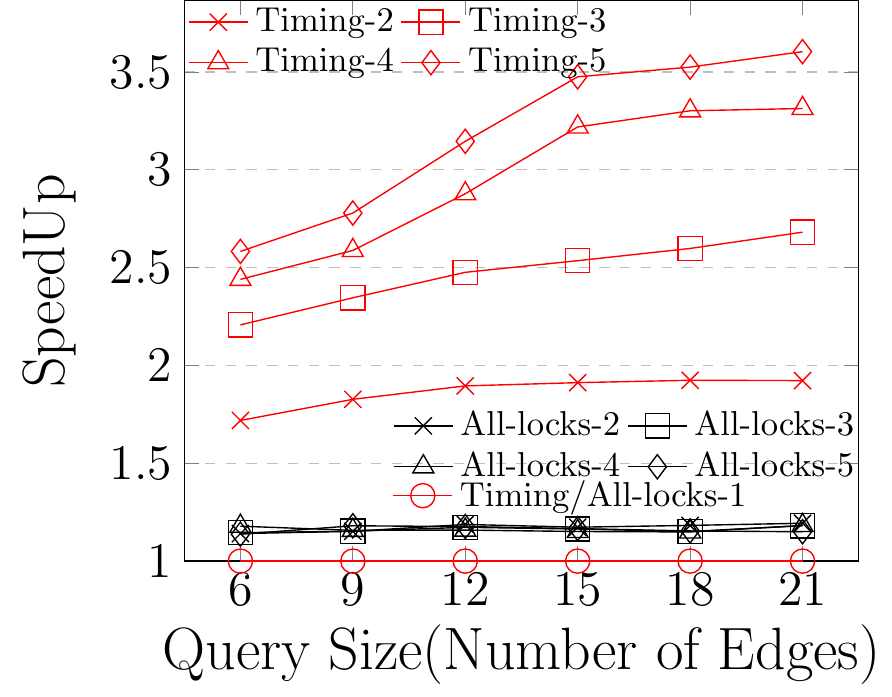}
         }
         \caption{Wiki-talk}
         \label{fig:spd:qsz:wkt}
     \end{subfigure}
     }
\caption{Speedup over Different Query Size}
\vspace{-0.15in}
\label{fig:spd:qsz}
\end{figure}

We evaluate the performance of our concurrency technique in this section by varying the number of threads running in parallel. We use \textbf{Timing-$N$} to differentiate different settings of parallel threads ($N$). We also implement, for comparison, a locking mechanism that requires a thread to obtain all locks before it is allowed to proceed (called \textbf{All-locks-$N$}). We present the speedup over single thread execution in Figures \ref{fig:spd:win}-\ref{fig:spd:qsz}. We can see that our locking strategy outperforms All-locks-$N$. As the number of threads grows, the speedup of our locking mechanism improves, while the speedup of All-locks-$N$ remains almost the same. 
\optionshow{}{
Our method applies the fine-grained locking strategy and different threads may not conflict. In this case, the degree of concurrency increases when the number of threads grows. However, All-locks-$N$ always locks all items that may be accessed in a thread. In other words, All-locks-$N$ is almost the same as the sequential algorithm. This is why the speed up of All-locks-$N$ is always about 1.2. 
} 
Figure \ref{fig:spd:qsz} also shows that speedup of our solution improves as the query size gets larger. In fact, the larger the query size, the more items tend to be in the corresponding expansion lists, which further reduces the possibility of contention.

\vspace{-0.1in}
\vspace{-0.1in}
\optionshow{
\vspace{-0.15in}
}{}

\begin{figure}[h!]
\centering
    \begin{subfigure}[t]{0.42\linewidth}
	    \centering
	    \resizebox{\linewidth}{!}
	    {
	        \includegraphics{\djfolder 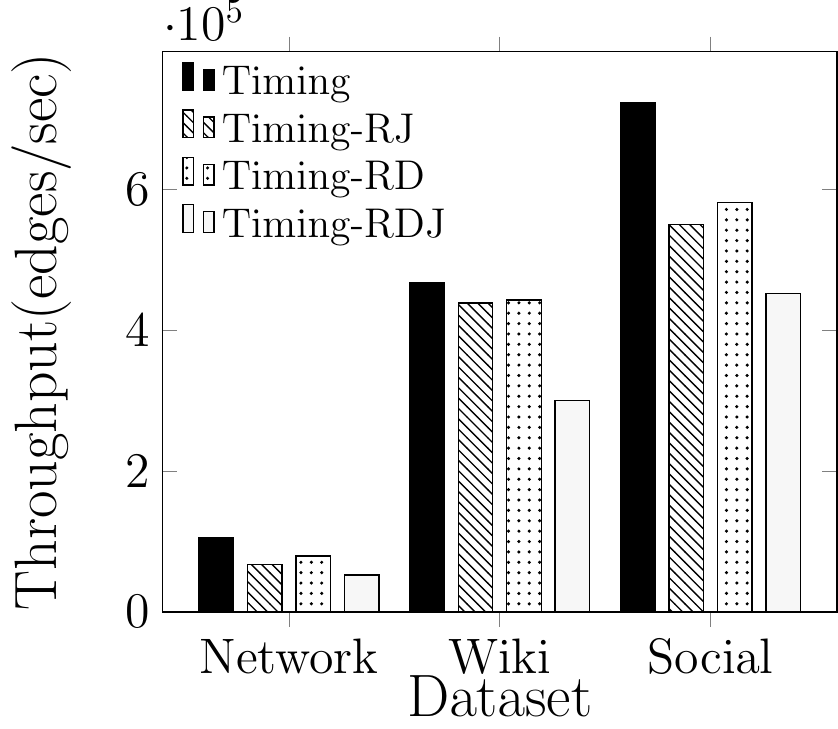}
	    }
	    \caption{Time Efficiency}
	    \label{fig:dj:time}
    \end{subfigure}
    \begin{subfigure}[t]{0.42\linewidth}
        \centering
        \resizebox{\linewidth}{!}
        {
            \includegraphics{\djfolder 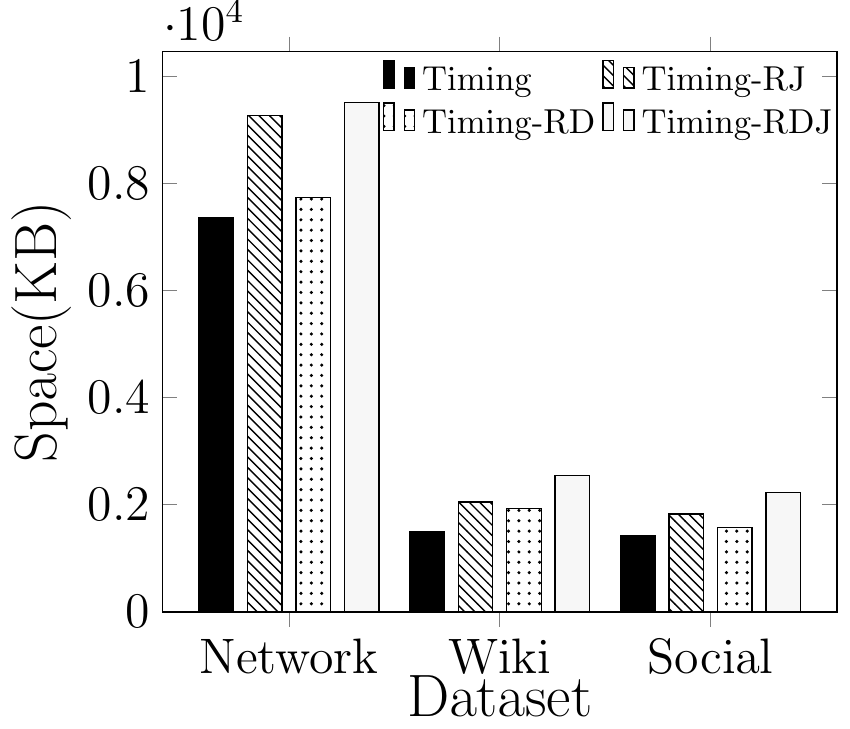}
        }
        \caption{Space Efficiency}
        \label{fig:dj:space}
    \end{subfigure}
\caption{Evaluating Optimizations}
\label{fig:dj}
\end{figure}
\optionshow{
\vspace{-0.25in}
}{}
\subsection{Decomposition and Join Order}
We evaluate the effectiveness of our decomposition strategy and selection of the join order. We implement three alternative solutions:  to evaluate the decomposition strategy, we design an alternative that randomly retrieves a decomposition from $TCsub(Q)$ for a given query $Q$ (denoted as \textbf{Timing-RD}); to evaluate the join order selection, we design a second alternative that randomly chooses a prefix-connected sequence (join order) over a given decomposition $D=$ $\{P_1$, $P_2$, ..., $P_k\}$ (denoted as \textbf{Timing-RJ}), and a third that applies random decomposition and uses random prefix-connected sequence (denoted as \textbf{Timing-RDJ}). In the evaluation, we fix the window size to $30,000$. Figure \ref{fig:dj} shows that our solution outperforms the alternatives. The main reason is that the decomposition  and join order strategy reduces the partial matches we need to maintain, which further helps reduce the time cost for computation over those partial matches.

\optionshow{}{
\subsection{\mred{Case Study}}

We evaluate our solution over an internal non-anonymous network traffic data. Note that we did not use the network traffic data from CAIDA since it is anonymous and there is no way for us to verify whether the detected patterns indicate attacks or not. Our collected dataset contains all traffic data of more than $20$ windows/linux servers/PCs. The time span of the traffic starts from Sept. 1st, 2017 while ends at Dec. 11th, 2017, when one windows server (of IP address 59.**.222.36) was found compromised as slaves of a ZeuS botnet. It is confirmed that the windows server was compromised at Nov. 28th, 2017 when it started to frequently communicate with a C\&C server (of IP address 101.*.81.189) through domain names that are created by Gameover DGA.


The pattern (query graph) we monitor is exactly the one in Figure \ref{fig:networktraffic}.
We set the window size of 30 seconds which is long enough for an attack of such pattern. We remove all traffic accessing top 10,000 websites in Alexa Rank\footnote{https://www.alexa.com/topsites}, which is a common source of whitelist of websites in cyber-security field.
We find that our algorithms successfully detected a match happened at Nov. 28th as presented in Figure \ref{res-fig:resultpattern}. The timestamp of each edge is in the original format when processed by tcpdump. Specifically, the window server ($V^1$) of IP 59.**.222.36 communicated with a web server (IP: 125.**.254.78 through website address ``www.***.edu.cn/27/ketiyanshou.htm'') and then built connection with a C\&C server (IP: 101.*.81.189 through domain name ``tevmwugbtzp8jychaelt1ggb.net'').  
Apparently, if we monitor this pattern on the Windows server at that time, we can  stop the connections to the C\&C server much sooner.


\begin{figure}[h!]
\centering
\resizebox{0.85\linewidth}{!}{
	\includegraphics{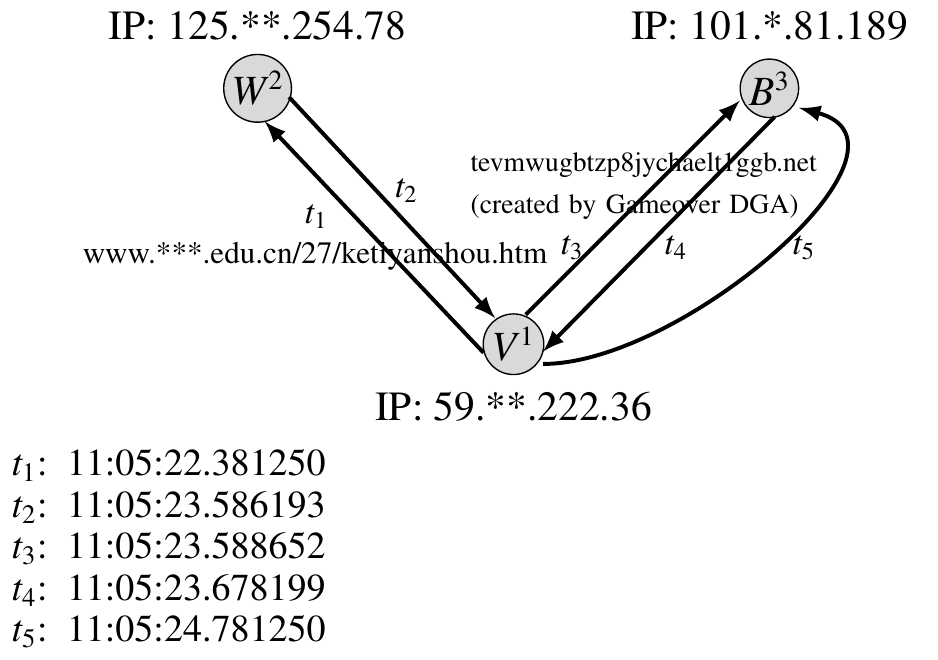}
}
\caption{Detected attack graph}
\label{res-fig:resultpattern}
\end{figure}

}


\optionshow{}{
\subsection{Decomposition Size $k$}\label{sec:exp:sizek}
We evaluate our solution with comparative ones over query of different decomposition size ($k$). We fix the query size as $12$ and window size as $30,000$. We set five different $k$: $1$, $3$, $6$, $9$, $12$. To generate a query of a specific decomposition size $k$, we can constantly create timing order $\prec$ over a retrieved subgraph $g$ (by varying permutation of $g$'s edges) until $g$ and $\prec$ constitute a query that can be decomposed into $k$ TC-subqueries according to our decomposition strategy. In fact, for $k=1$, we assign the timing order between every two edges in $g$ according to their timestamps in the data graph, while for $k=12$, we just set the timing order as $\emptyset$.  We present the throughput and space cost of all methods over different $k$ in Figure \ref{fig:time:ratio} and Figure \ref{fig:space:ratio}, respectively. Our method outperform existing ones obviously. Specifically, the throughput of our method is higher than others by nearly one order of magnitude and our data structure cost much less space than comparative ones do. With the decomposition size increases, the throughput of our method decrease while the space cost increase, which confirms our claim that $k$ should be as small as possible.

\begin{figure}[h!]
\centering
    \begin{subfigure}[t]{0.45\linewidth}
	    \centering
	    \resizebox{\linewidth}{!}
	    {
	        \includegraphics{\timefolder 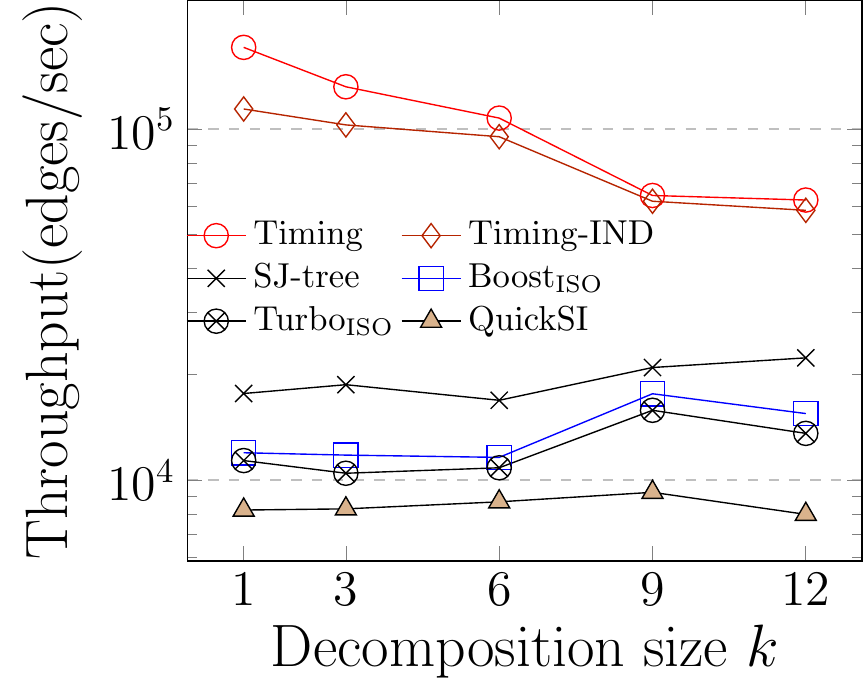}
	    }
	    \caption{Network Flow}
	    \label{fig:time:ratio:net}
    \end{subfigure}
      	\hspace{0.1in} 
    \begin{subfigure}[t]{0.45\linewidth}
        \centering
        \resizebox{\linewidth}{!}
        {
            \includegraphics{\timefolder 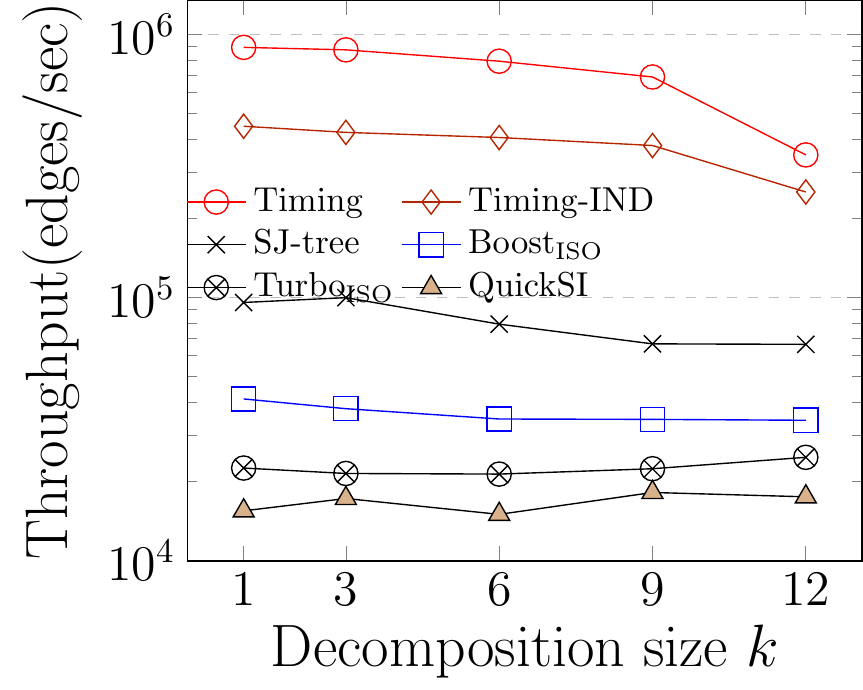}
        }
        \caption{Social Stream}
        \label{fig:time:ratio:rdf}
    \end{subfigure}
    \begin{subfigure}[t]{0.45\linewidth}
        \centering
        \resizebox{\linewidth}{!}
        {
            \includegraphics{\timefolder 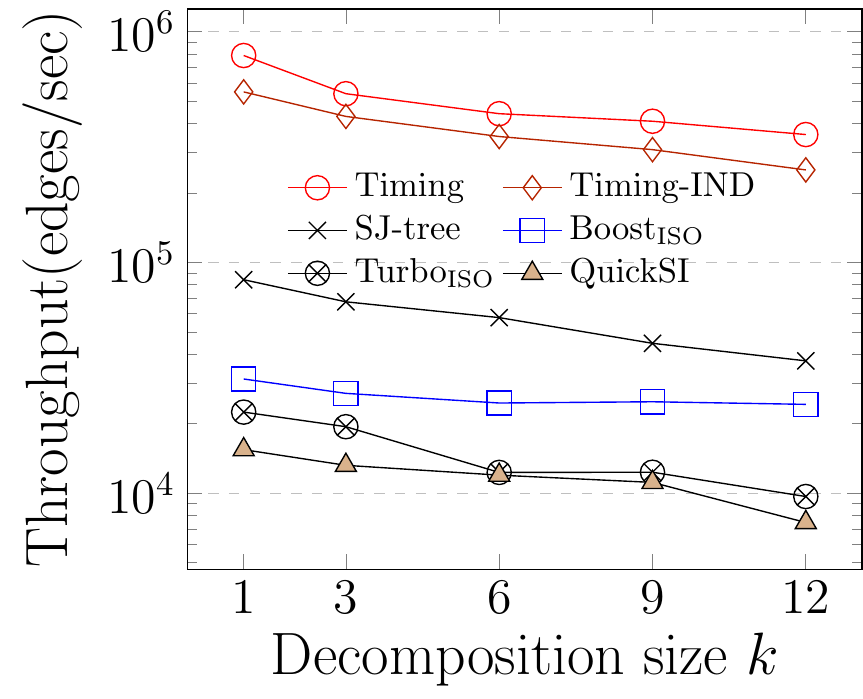}
        }
        \caption{Wiki-talk}
        \label{fig:time:ratio:wkt}
    \end{subfigure}
\caption{Throughput over Different $k$}
\label{fig:time:ratio}
\end{figure}
\begin{figure}[h!]
\centering
    \begin{subfigure}[t]{0.45\linewidth}
	    \centering
	    \resizebox{\linewidth}{!}
	    {
	        \includegraphics{\spacefolder 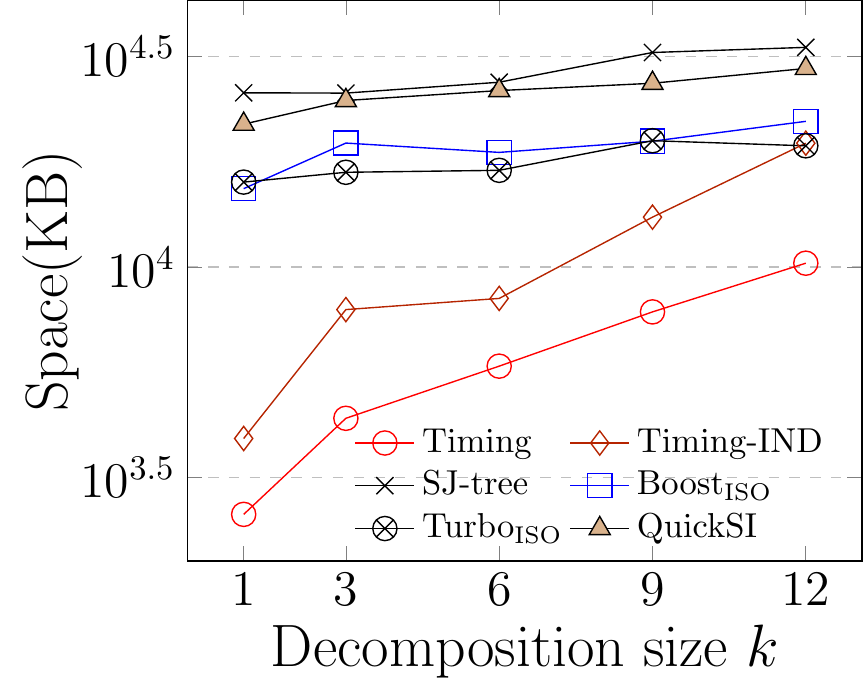}
	    }
	    \caption{Network Flow}
	    \label{fig:space:ratio:net}
    \end{subfigure}
      	\hspace{0.1in} 
    \begin{subfigure}[t]{0.45\linewidth}
        \centering
        \resizebox{\linewidth}{!}
        {
            \includegraphics{\spacefolder 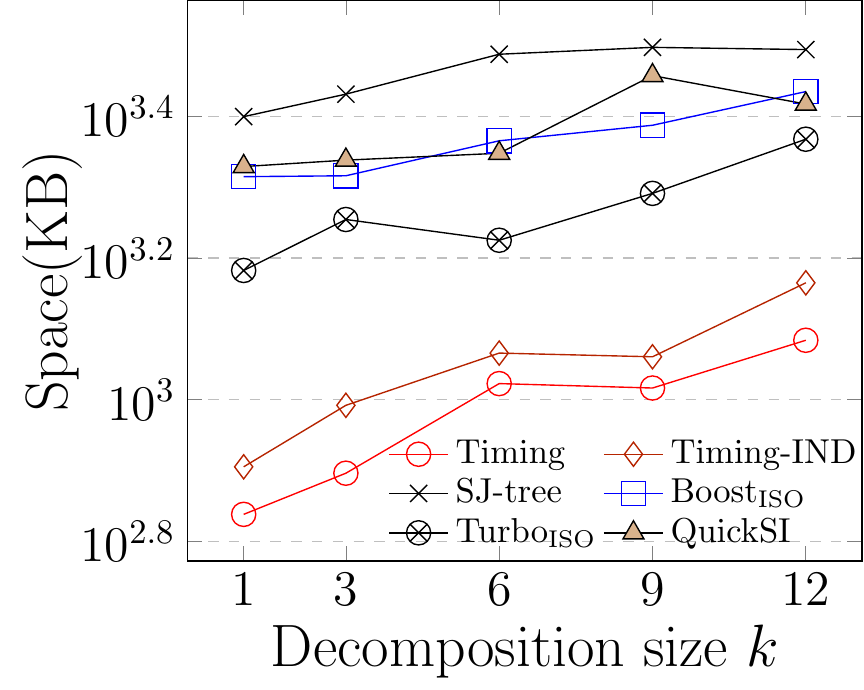}
        }
        \caption{Social Stream}
        \label{fig:space:ratio:rdf}
    \end{subfigure}
    \begin{subfigure}[t]{0.45\linewidth}
        \centering
        \resizebox{\linewidth}{!}
        {
            \includegraphics{\spacefolder 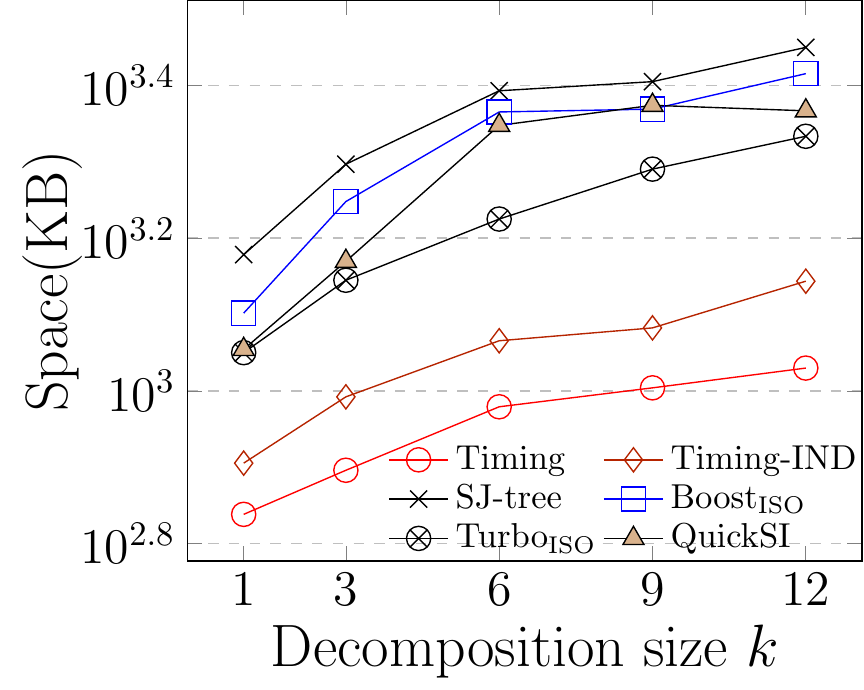}
        }
        \caption{Wiki-talk}
        \label{fig:space:ratio:wkt}
    \end{subfigure}
\caption{Space over Different $k$}
\label{fig:space:ratio}
\end{figure}
} 

\optionshow{}{
\subsection{Selectivity of Query Set}
\label{sec:appendix:selectivity}
We report the selectivity of the generated queries varying window size and query size. We present the corresponding average number of answers of these generated queries in Figure \ref{fig:selectivity}. We can see that the number of answers almost decreases with the growth of the query size while increases with the growth of the window size.

\begin{figure}[h!]
\centering
    \begin{subfigure}[t]{0.45\linewidth}
	    \centering
	    \resizebox{\linewidth}{!}
	    {
	        \includegraphics{\selectfolder 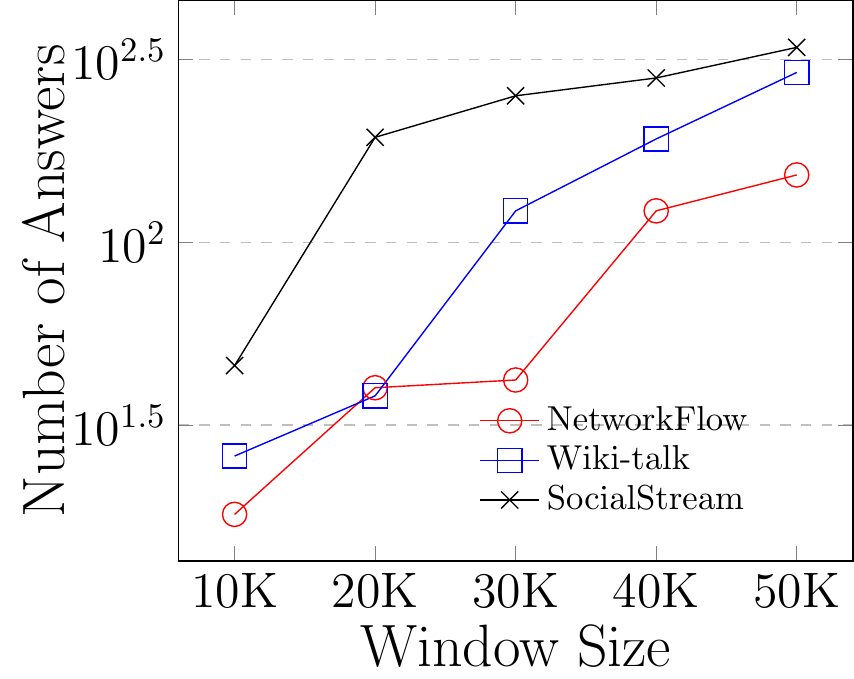}
	    }
	    \caption{Selectivity Varying Window Size}
	    \label{fig:selectivity:win}
    \end{subfigure}
      	\hspace{0.1in} 
    \begin{subfigure}[t]{0.45\linewidth}
        \centering
        \resizebox{\linewidth}{!}
        {
            \includegraphics{\selectfolder 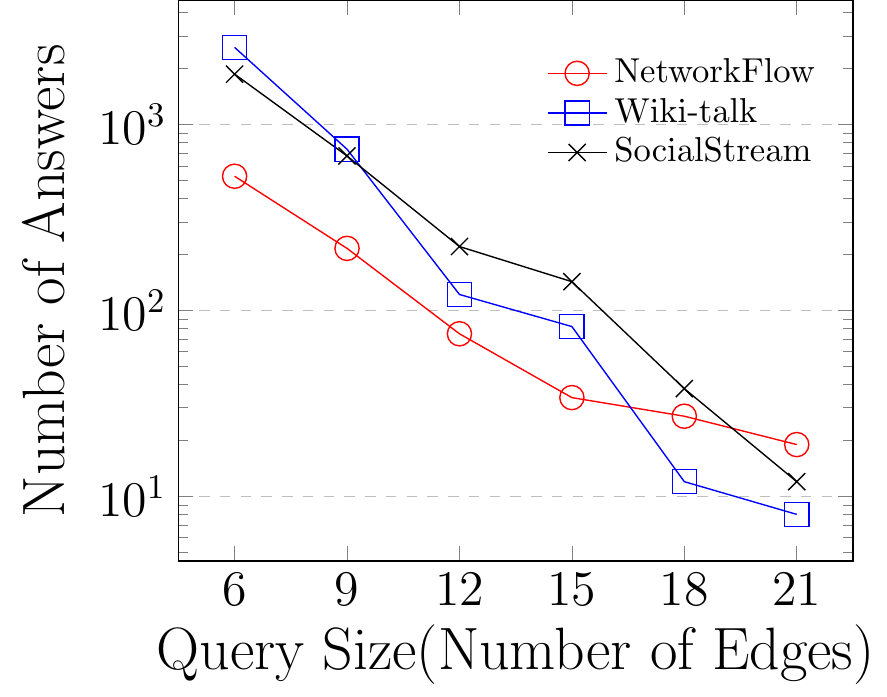}
        }
        \caption{Selectivity Varying Query Size}
        \label{fig:selectivity:qsz}
    \end{subfigure}
\caption{Selectivity}
\label{fig:selectivity}
\end{figure}
}

\section {Conclusions}
\label{sec:con}
The proliferation of high throughput, dynamic graph-structured data raises challenges for traditional graph data management techniques. This work studies subgraph isomorphism issues with the timing order constraint over high-speed streaming graphs. We propose an expansion list to efficiently answer subgraph search and propose MS-tree to greatly reduce the space cost. More importantly, we design effectively concurrency management in our computation to improve system's throughput. To the best of our knowledge, this is the first work that studies concurrency management on subgraph matching over streaming graphs. Finally, we evaluate our solution on both real and synthetic benchmark datasets. Extensive experimental results confirm the superiority of our approach compared with the state-of-the-arts subgraph match algorithms on streaming graphs.

\balance

\bibliographystyle{IEEEtran}
\bibliography{main}  


\end{document}